\documentstyle[12pt]{l-aa}
\input{psfig}

\def\Real{{\rm I\mathchoice{\kern-0.70mm}{\kern-0.70mm}{\kern-0.65mm}%
  {\kern-0.50mm}R}}

\font \bolditalics = cmmib10
\def\bx#1{\leavevmode\thinspace\hbox{\vrule\vtop{\vbox{\hrule\kern1pt
        \hbox{\vphantom{\tt/}\thinspace{\bf#1}\thinspace}}
      \kern1pt\hrule}\vrule}\thinspace}

\def \vc #1{{\textfont1=\bolditalics \hbox{$\bf#1$}}}

\def\rmk{{\rm k}}

\def\qg{{\bf q}}

\def\kg{{\bf k}}

\def\xig{{\vc \xi}}
\def\xigs{{\vc \xi^S}}
\def\xigi{{\vc \xi^I}}
\def\varphig{{\vc \varphi}}
\def\thetag{{\vc \theta}}

\def\Dg{{\vc D}}

\def\Dc{{\cal D}}

\def\be{\begin{equation}}
\def\ee{\end{equation}}
\def\ba{\begin{eqnarray}}
\def\ea{\end{eqnarray}}
\def\mg{\big{<}}
\def\md{\big{>}}

\def\d{{\rm d}}
\def\omb{{\overline{\omega}}}

\begin{document}

%\input{psfig.sty}
%\input{epsf.sty}
%\input{macros.tex}
% \input macros-nofig

%\bigskip
   \thesaurus{02         % A&A Section 2: Cosmology
              (12.03.4; 
               12.04.1;  
               12.07.1;  
               12.12.1)} 

%   \title{Large Scale Structure Mass Reconstruction from Weak Lensing}
%  \title{Noise effects on precision measurements in weak lensing survey}

   \title{Efficiency of weak lensing surveys to probe cosmological models}
   \author{L. van Waerbeke$^{1,2,3}$, F. Bernardeau$^4$,Y. Mellier$^{5,6}$}
   \offprints{waerbeke@cita.utoronto.ca}

  \institute{$^1$ CITA, 60 St Georges Str., Toronto, M5S 3H8 Ontario, Canada.\\
   $^2$ MPA, Karl-Schwarzschild-Str. 1, Postfach 1523, 
D-85740 Garching, Germany. \\
   $^3$ OMP, 14 av. Edouard Belin, 31400, Toulouse, France.\\
   $^4$ Service de Physique Th\'eorique. C.E. de Saclay. 91191
Gif-sur-Yvette Cedex, France. \\
   $^5$ Institut d'Astrophysique de Paris. 98 bis, boulevard
Arago. 75014 Paris, France. \\
   $^6$ Observatoire de Paris. DEMIRM. 61, avenue de
l'Observatoire.  75014 Paris, France.}

   \markboth{Efficiency of weak lensing surveys}{L. van Waerbeke et al.}

%\date{Received, , Accepted, }

\maketitle
   \markboth{Efficiency of weak lensing surveys}{L. van Waerbeke et al.}

\begin{abstract}
We apply a mass reconstruction technique to
large-scale structure gravitational distortion maps, simulated for different
cosmological scenarii on scales from $2.5$ arcmin to $10$
degrees.
The projected mass is reconstructed using a non-parametric least square
method involving the reduced shear on which noise due to intrinsic galaxy
ellipticities  has
been added. The distortion of the galaxies is calculated using the 
full lens equation, without any
hypothesis like the weak lensing approximation, or other linearization.

It is shown that the noise in the reconstructed maps is perfectly
uncorrelated Poissonian, with no propagation from short to large scales. The
measured power spectrum and first four moments
of the convergence can be corrected accurately for this source of
noise. The cosmic variance
of these quantities is then analyzed with respect to the density of
the background galaxies using 60 realizations of each model.
We show that a moderately deep weak lensing survey
($5\times 5$ degrees with a typical background population of
$30~$gal/arcmin$^2$ at a redshift $z_s\simeq 1$)
is able to probe the amplitude of the power spectrum with 
a few percent accuracy for models with $\sigma_8\ \Omega^{0.8}=0.6$.

Remarkably, we have found that, 
using the third moment of the local convergence only, 
such a survey would lead to a $6~\sigma$
separation between open ($\Omega=0.3$) and flat ($\Omega=1$) models.
This separation does not require a very deep survey, and it
is shown to be robust
against different hypothesis for the normalization or the shape
of the power spectrum. 

Finally, the
observational strategy for an optimal measurement of the power spectrum
and the moments of the convergence is discussed. 

\end{abstract}

%\vfill\eject

\section{Introduction}
Mass reconstruction from gravitational distortion inversion is 
a promising technique to probe the
mass distribution and the clustering on very large scales, regardless of
the nature and the dynamical state of the dark and luminous matter.
Pioneering theoretical work done by Gunn (1967), Jaroszy\'nski et al. (1990),
Blandford et al. (1991), Miralda-Escud\'e
(1991) and Kaiser (1992) has shown that the expected distortion amplitude
of weak lensing effects produced by large scale 
mass fluctuations ($\ge $ 1Mpc) 
is roughly at the percent level. This low level of distortion is
observable due
to the large number density of galaxies observed in deep surveys (Kaiser 1992). 
Needless is to say that the 
observation of such distortion fields provides a unique way to build
a picture of the large scale mass distribution, independent on any
biasing and/or dynamical prescriptions.

The scientific impact
on the determination of cosmological parameters from weak lensing
surveys has been underlined by recent theoretical work. Most of the
papers quoted above show that the distortion two-point correlation
function can be used to constrain the mass power spectrum. Villumsen
(1996) remarked that the amplitude of the local distortion is proportional
to the amplitude of the 3D density fluctuations {\sl and} roughly to the
density parameter $\Omega_0$. Bernardeau et al. (1997, hereafter BvWM)
extended these
calculations in the $\Omega_0-\lambda$ plane and show that the amplitude
is also slightly dependent on the cosmological parameter $\lambda$.
In the same paper the authors also show that the shape of the
convergence probability distribution function
can be used to disentangle the $\Omega$ and $\sigma_8$ dependence
for models of large-scale structure formation with Gaussian initial
conditions. In particular they demonstrate that the skewness\footnote{results
are explicitly given at beginning of Sect. 4}, third moment
expressed in terms of the square of the second, is
roughly inversely proportional to the total mass density
of the Universe $\Omega_0$,
but independent on the amplitude of the fluctuations, as well as the shape
of the power spectrum.
This result, obtained by means of perturbation theory, is expected
to be exact at large enough scale (see Gazta\~naga \& Bernardeau, 1998).
The skewness
should be enhanced in the nonlinear regime at small scales
(Gazta\~naga \& Bernardeau 1998, see also Colombi et al. 1997, and
Jain et al. 1998) which amplifies the
differences between open and flat cosmologies.
So, even if definitive quantitative predictions for such
a quantity cannot be given from our present knowledge, it is clear that 
the skewness can accurately discriminate between different cosmological models.

%These theoretical investigations have definitely  established
%the fact that 
%weak lensing surveys can be used to determine the power spectrum
%shape and amplitude, and the density parameter $\Omega_0$, provided that
%reliable convergence maps at the percent level can be obtained from the
%observations.

Recently many theoretical
and observational aspects have been investigated in detail in order
to converge towards an unbiased measurement of such small distortions
(Bonnet \& Mellier 1995, Kaiser et al. 1995, and Van Waerbeke et al. 1997). 
The technical limitations that have been recognized so far
come from systematics due to the correction of spurious distortions
generated by the optic defects, fuzzy-shaped point spread function (PSF), and pixel convolution (sampling).
Available image analysis techniques and image
quality ensure that such systematics can now be reduced to a one
percent level, which is necessary for these weak lensing applications,
but beyond the need for cluster mass reconstruction.
The ultimate limitation for the use of weak lensing surveys as a cosmological
probe depends therefore on the accuracy with which the correction of 
the spurious distortions can be corrected. On the other hand, 
if the systematics can be reduced
to the sub-percent level, the weak lensing analyses are limited
by the intrinsic ellipticities of galaxies which acts as a shot
noise for the gravitational distortion effect.

In this paper we investigate the effects of the shot noise and finite size
survey (cosmic variance) on the
determination of the cosmological parameters and the power
spectrum. In particular, two goals are looked for,
\begin{itemize}
\item the precision on the determination of some cosmological parameters
that can be expected from such maps;
\item the maximum level of observational systematics 
which is acceptable in order to achieve these theoretical precisions.
\end{itemize}

This study is made in particular in the perspective of 
ongoing and future wide field deep imaging surveys 
devoted to 
weak lensing analysis (for instance the MEGACAM project, Boulade et al. 1998)
at CFHT  \footnote{See also
the WEB page http://terapix.iap.fr/terapix\_megacam.html.}
or the SDSS project, Stebbins 1996). In such a perspective,
there are still open issues concerning the optimal observing strategy that
should be adopted:
\begin{itemize}
\item Wide and shallow survey versus deep and narrow.
\item The determination of the optimal shape and size of the survey
depends on the noise properties, and on the correlation properties
of the signal. The investigations that have been done so far
(Kaiser 1992, 1998 and Seljak 1997) assume that the projected mass
follows a Gaussian statistics. If such an assumption had to be dropped it
may significantly change the conclusions that have been reached.
\end{itemize}

In addition, since the reconstruction of the projected mass from the shape of
the galaxies is neither local nor linear in terms of the distortion
field (Kaiser 1995) and of the intrinsic ellipticities of the galaxies,
it is essential to understand how the noise
propagates in the reconstructed mass maps. In order to investigate
how these different effects may couple together, we built a series of 
projected mass
maps that contain a realistic amount of non-Gaussianity. The
associated distortion field is then derived on the basis
of the full non-linear lens equation. A noise is added on the distortion
maps, and the convergence is finally reconstructed. This paper presents the
statistical analysis of those reconstructed maps for different cosmological
models and different observational contexts.

The details of mass map generation and useful definitions are presented
in Sect. 2.
Sect. 3 presents the power spectrum analysis, the noise properties
in the reconstructed mass maps, and the cosmic variance on the estimated
power spectrum. Sect. 4
repeats similar analysis on moments in real space, where a comparison
between top hat and compensated filter is done. It contains also
some highlights about other possible statistical quantities
that could be used to measure $\Omega$, and a
comparison with results obtained for different power spectra.
We finally summarize our results and discuss the best observational
strategies.

\section{Generation of realistic $\kappa$-maps}

\subsection{Lensing effects, displacement and amplification matrix}

Any mass concentration
deflects light beams by an angle
proportional to the gradient of the local gravitational potential.
This effect induces an apparent displacement of the sources,
so that a source that was at the angular position $\xigs$ will be observed
at position $\xigi$, with
\be
\xigs=\xigi-{2\over c^2}{\Dc_K(\chi_s-\chi)\over \Dc_K(\chi)
\Dc_K(\chi_s)}\int\d \chi\ \nabla\phi(\chi,\xigi),\label{disp}
\ee
where $\chi$ is the radial coordinate ($\chi_s$ is the one of the source), 
$\Dc_K(\chi)$ the co-moving angular
diameter distance and $\phi(\chi,\xig)$ the 3D gravitational
potential. The differential displacement of the images induces
an image distortion, which depends on the second derivatives of the
gravitational potential, i.e. on the mass density and the tidal field.
This gravitational lensing effect of a thin lens is therefore characterized
by an isotropic stretching, described by the 
convergence $\kappa$, and an anisotropic distortion given by the complex 
shear $\gamma=\gamma_1+i\gamma_2$.
The so-called amplification matrix $\cal A$ describes the change of 
local coordinates between the source and image planes and can be written as
\begin{equation}
{\cal A}=\pmatrix{1-\kappa-\gamma_1& -\gamma_2 \cr -\gamma_2 & 1-\kappa+\gamma_1}.
\end{equation}
Its elements are related to the first derivative of the displacement field,
\ba
\kappa&=&{1\over 2}\ (2-\nabla\cdot\xigs)\\
\gamma_1&=&{1\over 2}
\left(\nabla_{2}\xi_2^S-\nabla_{1}\xi_1^S\right) \\
\gamma_2&=&-\nabla_{1}\xi_2^S,
\ea
and are therefore related to the projected gravitational potential 
of the lens $\psi$ via,
\begin{eqnarray}
\kappa=\Delta \psi/2; \ \ 
\gamma_1=\left(\psi_{11}-\psi_{22}\right)/2;\ \ 
\gamma_2=\psi_{12}.
\label{lens_def}
\end{eqnarray}
and where $\psi$ is given by,
\begin{equation}
\psi(\varphig,\chi_s)={2\over c^2} \int_0^{\chi_s}~d\chi'{\Dc_K(\chi-\chi')\over \Dc_K(\chi)\Dc_K(\chi')}
\phi(\Dc_K(\chi')\varphig,\chi'),
\label{psi_def}
\end{equation}
where $\varphig$ is the angle such that $\xig=\Dc_K(\chi)\varphig$.
Here $\psi$ depends on $\chi_s$ since the potential is integrated
from a given source plane $\chi_s$ to the observer. This can be trivially
generalized to a source redshift distribution. This equation is valid
only if the lens-lens coupling is dropped and the Born approximation is
used. Th validity of this assumption has been discussed by
BvWM and Schneider et al. 1997 (hereafter SvWJK).

\subsection{The galaxy shape matrices}

A source galaxy may be described by a complex ellipticity defined as
\begin{equation}
\epsilon^{(s)}={1-r\over 1+r}e^{2i\vartheta}
\end{equation}
where $\vartheta$ is the galaxy orientation, and $r$ is the square root
of the ratio of the eigenvalues of the shape
matrix $Q^{(s)}_{ij}$ of the (centered) surface brightness profile of
the galaxy,
\begin{equation}
Q^{(s)}_{ij}={\int \d\thetag~\theta_i\theta_j~{\cal S}(\thetag)\big/ 
\int \d\thetag{\cal S}(\thetag)               %\theta^2/2
}.
\label{moments_def}
\end{equation}
In presence of lensing, the shape matrix of the image of the galaxy is given by 
$Q={\cal A}^{-1}Q^{(s)}{\cal A}^{-1}$, and provided that the amplification
matrix does not vary over the galaxy's area, the observed ellipticity
is still described by Eq. (\ref{moments_def}), and it writes
(Schneider \& Seitz 1995),
\begin{equation}
\epsilon={\epsilon^{(s)} +g\over 1+g^\star\epsilon^{(s)}},
\label{elli_single}
\end{equation}
where $g=\gamma/(1-\kappa)$ is the complex reduced shear. The observable is the distortion
$\delta=2g/(1+|g|^2)$, but for sub-critical lenses 
like those we discuss in this work, $g$ is also directly observable,
\begin{equation}
|g|=\sqrt{1/|\delta|-1}-1/|\delta|.
\end{equation}
If the orientation of the source galaxies is random
($\langle \epsilon^{(s)}\rangle=0$), the observed
mean ellipticity of galaxies is an unbiased estimate of the reduced
shear (Schramm \& Kayser 1995, Schneider \& Seitz 1995),
\begin{equation}
\langle \epsilon\rangle=g.
\label{local_g}
\end{equation}
The weak lensing approximation ($\langle \epsilon\rangle\simeq \gamma$)
is generally used in the case of lensing by large scale structures.
However, we will not use that since we want to analyze the noise
propagation in mass map reconstructed in the full non-linear regime
(Eq. (\ref{elli_single})).

\subsection{Construction of the synthetic projected mass maps, and their
reconstruction}

The adopted procedure to build and analyze projected density maps is: \\

1) Generation of a density map, which directly provides the convergence
$\kappa$ given the source redshifts:

In order to make a precise analysis of the cosmic
variance and noise properties in the reconstructed mass maps
it is necessary to have a large set of simulations of large scale
structure. Instead of CPU intensive $N$-body codes we use numerical
fast second order Lagrangian dynamics
(Moutarde et al. 1992) which has been shown to accurately reproduce the
statistical properties of LSS (Munshi et al. 1994,
Bernardeau et al. 1994, Bouchet et al. 1995). This allow us to build a
large amount of projected mass maps for different cosmological models.
%and to compare the statistics of $\kappa$
%for low and high density Universes. To
%this end, the mass fluctuations are normalized to $\langle \kappa^2\rangle$
%rather than $\langle \delta^2\rangle$, (where
%$\delta$ would be the 3D density contrast).
The appendix A gives details
about the generation of these maps and compares the
non-Gaussian features with expectations from real dynamics.

2) Calculation of the associated reduced shear map $g$ and addition of a
given level of noise to the reduced shear map:

From the previous projected mass maps, a gravitational distortion map is
computed, and the noise due to intrinsic ellipticities is added,
as described in Appendix B.
A "high" and "low" noise levels are simultaneously considered which correspond
respectively to a mean number density of galaxies of $\bar
n=30~$gal/arcmin$^2$ and $\bar n=50~$gal/arcmin$^2$. For the I-band, these
number densities are reachable with respectively $1.5$ or $4.5$ hours exposure 
at CFHT which is expected to provide galaxies of redshift
of about unity.
At this stage we have at our disposal 
a large number of maps that are supposed to mimic accurately
what can be observed with large CCD cameras.

3) Reconstruction of the original $\kappa$ map from $\epsilon$:

Many mass reconstruction methods already exists
(Seitz et al. 1998, Squires \& Kaiser, 1996, and references
therein). In this work, we use a non-parametric least square %$\chi^2$
method (see Bartelmann et al. 1996 for details) which
overfits data if no regularization process is used. The
hope by doing so is to preserve all the
noise properties intact, so that a detailed noise analysis can be done.
%The shear field is discretized over a regular grid, and
%the $\kappa$-field is reconstructed by minimizing the function,
%\begin{equation}
%\chi^2={\displaystyle \sum_{ij}} 
%\left|g(\psi_{ij})-\langle \epsilon\rangle\right|^2,
%\label{chi2}
%\end{equation}
%where the sum is performed over gridpoint coordinates $(i,j)$.
%The reduced shear $g(\psi)$ is calculated from the
%unknown potential $\psi$. It is chosen such to predict the mean observed
%ellipticities as good as possible.
The appendix B gives more technical details about the reconstruction algorithm.

\begin{figure*}
\centerline{
\psfig{figure=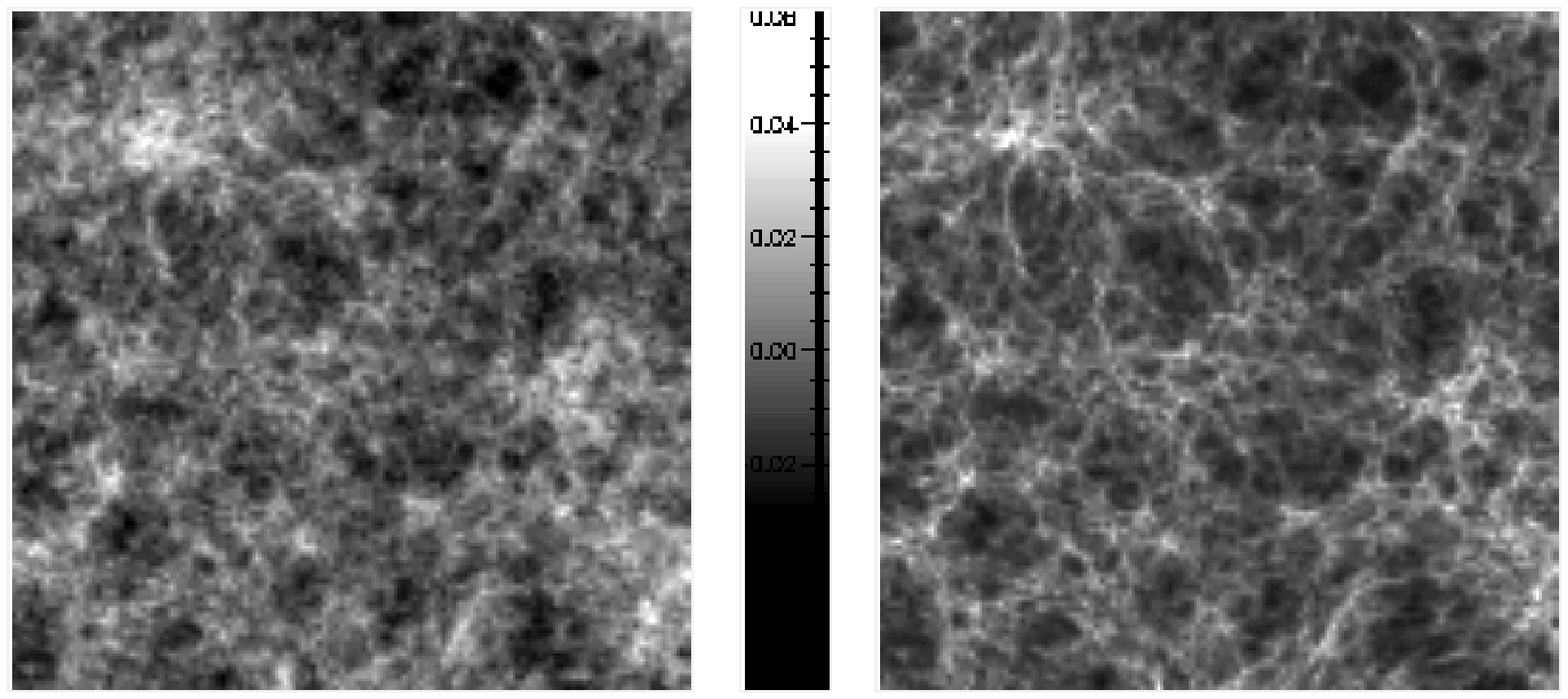,height=7cm}}
\centerline{
\psfig{figure=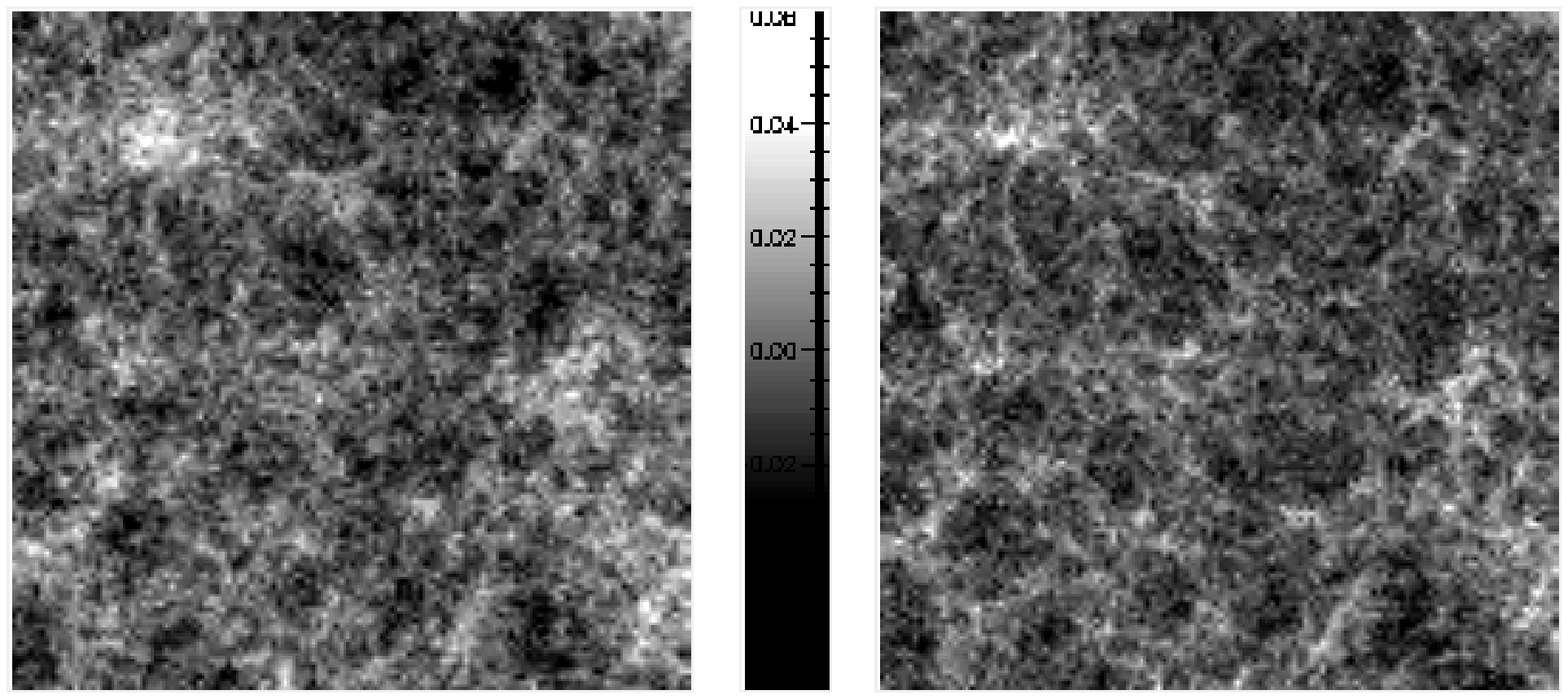,height=7cm}}
\caption{\label{kappa_model.ps} Example of reconstructions of
projected mass maps. The top panels show the initial noise-free
$\kappa$ map for either $\Omega=1$ (left panel) or $\Omega=0.3$ (right panel)
with the same underlying linear random field (see Appendix A) and
the same rms distortion.
The bottom panels show the reconstructed $\kappa$ maps with noise included in
the shear maps.
The maps cover a total area of  25 degrees$^2$. Each pixel has an
angular size of
2.5 arcmin$^2$ and averages the shear signal expected from deep CCD exposures
(about 30 galaxy/arcmin$^2$). The sources are assumed 
to be all at redshift unity and to have an intrinsic ellipticity distribution 
 given by Eq. (\ref{elli_def}). Such a survey is easily accessible to
MEGACAM at CFHT.
The precision with which the images can be reconstructed and the
striking differences between the two cosmological models
demonstrate the great interest such a survey would have.
}
\end{figure*}

Most of our generated images have an angular size of $5\times5$ degrees,
$120\times120$ pixels, each pixel having a $2.5'$ angular side 
(hereafter, the {\it superpixel} size) \footnote{The reason for that
pixelisation is
that second order Lagrangian dynamics is not able to probe high density
peaks (i.e. small fields.}. These are
typical scales that a MEGACAM survey at CFHT could probe.
Images of size $10\times 10$ degrees
($240\times 240$ pixels) have also been generated, in order to estimate how
the cosmic variance depends on the survey size in the non-linear regime, and to
compare the merit of deep-small survey area versus shallow-large survey.
For each cosmological model and observational context series of $60$ compact
maps are produced. Table 1 summarizes the cases that investigated.

\begin{table}
\caption{List of simulations which have been carried out. The power
spectrum BG corresponds to the formula (\ref{BGpower}). It is the same
for $\Omega_0=1$ and $\Omega_0=0.3$. The CDM model
corresponds to a standard CDM with $\Gamma=0.5$. $\omb$ is the ratio
between the local convergence and the projected normalized
density contrast, $\kappa=\omb\ \delta_{2D}$ (see appendix A).
}
\label{TSimuls}
\begin{center}
\begin{tabular}{c|c|c|c|c|c}
\hline
Spectrum&$\Omega_0$&$\sigma_8$&$z_{\rm sources}$&$\omb$&size (deg$^2$)\\
\hline
BG&1.0&0.6&1.0&0.115&$5\times5$\\
BG&0.3&1.53&1.0&0.045&$5\times5$\\
BG&1.0&0.6&1.5&0.195&$5\times5$\\
BG&0.3&1.40&1.5&0.0837&$5\times5$\\
BG&1.0&0.6&1.0&0.115&$10\times10$\\
BG&0.3&1.53&1.0&0.045&$10\times10$\\
BG&1.0&1.0&1.0&0.115&$5\times5$\\
sCDM&1.0&0.6&1.0&0.115&$5\times5$\\
\hline
\end{tabular}
\end{center}
\end{table}

Fig. \ref{kappa_model.ps} shows two examples of initial $\kappa$ maps, and
the reconstructed mass of the noisy distortion maps. This panel
illustrates what MEGACAM should be able to get during only
5 nights (!): $25$ exposures in
the I band, $1.5$ hour each. A single MEGACAM field corresponds to
a size of $24 \times 24$ pixels on Fig. \ref{kappa_model.ps}, which is
clearly the required minimum area to detect large scale
structure features like super clusters, filamentary structures or voids.

\section{Power spectrum analysis of the reconstructed maps}
\subsection{Noise statistical properties of the reconstructed maps}

It is clear that the mass reconstruction process
does not produce any boundary effects (which is settled by definition). The
only boundary effect, slightly
detectable on the figure, is a larger level of noise at the edge of the field, due
to the change in the finite difference scheme at that position 
(see Appendix B).
The noise due to the intrinsic ellipticities of the galaxies is
clearly visible at small scales. 

Since the least $\chi^2$ method used to reconstruct the convergence is a local
process, it is unlikely that noise propagates on scales larger than
the pixel size \footnote{This is less evident in the case of non-local
mass reconstruction.}.

Fig. \ref{pdek_5x5.ps} shows the power spectrum analysis of $60$
reconstructed mass maps in the case of two
different cosmological models $\Omega=1$ (cases (a) and (b))
and $\Omega=0.3$ (cases (c) and (d)). Fig.
\ref{pdek_5x5.ps} (a) and (c) show the noise free power spectrum (solid
lines) the power spectrum measured on the reconstructed maps with
$\bar n=30~$gal/arcmin$^2$ (dotted lines) and for $\bar n=50~$gal/arcmin$^2$
(dashed lines). The plateau for the latter two cases is the
consequence of the intrinsic ellipticities of the galaxies: at small scales,
the power is dominated by the ellipticity of the galaxies, thus $P(\rm k)$
tends to be constant, generally much higher than the signal.
Figs. (b) and (d) show the difference of the power
spectrum measurements on the reconstructed maps with the noise-free power
spectrum; thin dotted line is for $\bar n=50~$gal/arcmin$^2$ and thin dashed
line for $\bar n=30~$gal/arcmin$^2$. For visibility, error bars for scales
larger than $20~$arcmin have been dropped. 

In 3D space, these angular scales correspond approximately
to scales from 1 to 30 $h^{-1}$Mpc.
We leave for later studies the problem of inverting the measured
projected $P(\rm k)$ to the 3D one. This aspect has already been explored
at large angular scale by Seljak (1997).

\begin{figure}
\centerline{
\psfig{figure=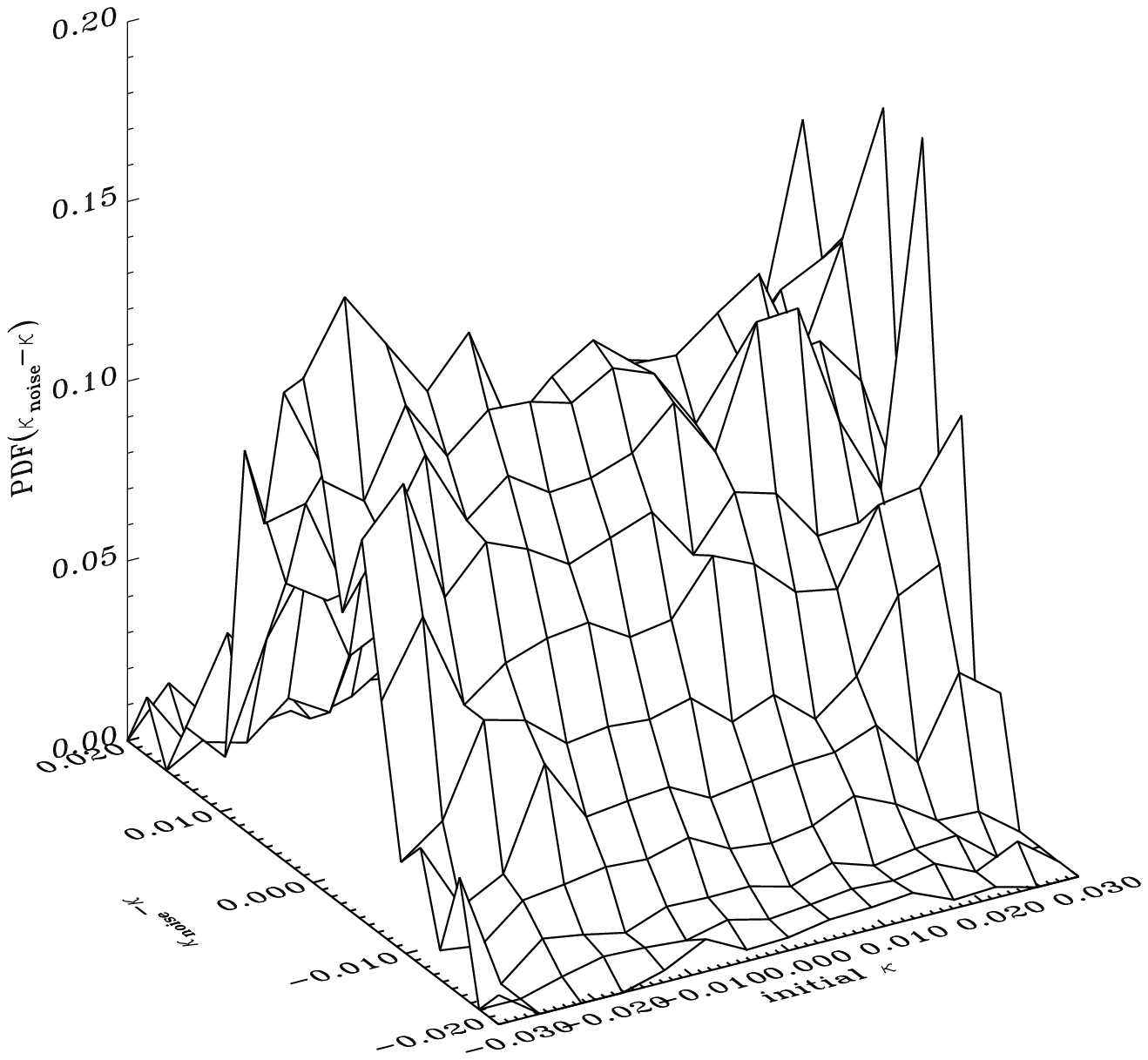,width=9cm}}
\caption{\label{noise.pattern} 
Histograms of the difference between the noisy mass map and the
initial mass map. The values of $\kappa$ have been selected
in bins. The histograms that are found to be all compatible with a
Gaussian distribution with a fixed mean and width.
}
\end{figure}

As can be inferred from the power spectra in Figure \ref{pdek_5x5.ps},
the noise has a flat power spectrum, characteristic of a white noise process.
Fig. (\ref{noise.pattern}) illustrates the fact that the
noise is independent on the underlying $\kappa$ field and follows a Gaussian
distribution.
The noise model introduced by Kaiser (1998) based on the weak
lensing approximation is now compared to the noise found in our simulations.
The weak lensing approximation applied to
Eq. (\ref{elli_single}) gives a local shear estimate
$\hat\gamma=\gamma+\bar \epsilon^{(s)}$, where $\gamma$ is the
true shear and $\bar \epsilon^{(s)}={\displaystyle 1/N_p\sum \epsilon^{(s)}}$,
the mean intrinsic ellipticities of $N_p$ galaxies in the superpixel $p$. 
Since the noise components are assumed to be spatially uncorrelated, the
statistical properties of the noise are,
\begin{equation}
\langle
{\bar \epsilon^{(s)}}_\alpha(\thetag_i){\bar \epsilon^{(s)}}_\beta(\thetag_j)\rangle=\sigma^2_\epsilon\delta^K_{\alpha\beta} \delta^K_{ij},
\label{noise_correl}
\end{equation}
where $\alpha,\beta=(1,2)$, $\delta^K$ is the Kronecker symbol and
$\sigma_\epsilon$ is the variance of one component of the intrinsic
ellipticities in one superpixel. The shear and the intrinsic ellipticities
of the galaxies are
uncorrelated in the weak lensing approximation.
The measured power (on the noisy mass maps) can then be expressed
only in terms of the
true power $P_\kappa(\rmk)$ (the one we want to estimate) and the
power spectrum of the noise (Eq.(\ref{noise_correl})). If $\tilde \kappa$
denotes the Fourier transform of the measured convergence, its power
spectrum is given by
\begin{equation}
\langle \tilde\kappa^2(\rmk)\rangle\simeq \bar n
P_\kappa(\rmk)+\bar\sigma^2_\epsilon.
\label{power_estim}
\end{equation}
This equation is only valid for a compact survey, where $\bar n$ is the
mean number density of the galaxies per superpixel, and $\bar\sigma_\epsilon$
is the
mean value of $\sigma_\epsilon$ over the survey. For a sparse survey, the first
term in Eq. (\ref{power_estim}) is changed into a convolution term, but the
noise contribution to the observed power spectrum remains independent on that
power spectrum. A convenient way to estimate $\bar\sigma^2_\epsilon$ is
to take 
\begin{equation}
{\displaystyle \bar\sigma^2_\epsilon\simeq {1\over n_{\rm pix}}
\sum_p\left({1\over N_p}\sum_{\rm gal}
\left[\epsilon^{(s)}_1\right]^2\right)},
\end{equation}
where $n_{\rm pix}$ is the number of superpixels and 
$N_p$ the number of galaxies
in the superpixel $p$. This is
nothing else but the variance of the observed ellipticities
of the selected galaxies\footnote{The noise depends only
on the variance, not on higher moments of the intrinsic ellipticity 
distribution}.

The thin dashed and dotted straight lines on Fig. \ref{pdek_5x5.ps} 
correspond to the expected noise power spectra (for $\bar n=30~$gal
/arcmin$^2$ and $\bar n=50$~gal/arcmin$^2$) according to the Kaiser's model
given by Eq. (\ref{power_estim}). It perfectly
fits the noise part of the mass reconstructed with noisy data, whatever the
cosmological model and the noise level. 
This is true even for the open cosmological models for
which stronger non-linearities could have produced a stronger coupling with
the noise. 
The fact that the noise component is pure white noise with an amplitude
in agreement with the theoretical prediction is a
remarkable result since the full non-linear equations were used, and
it shows that the weak lensing approximation can be safely
used to remove the noise component and to get an unbiased estimate of the
power spectrum, down to the smallest scales considered here.

\begin{figure*}
\centerline{
\psfig{figure=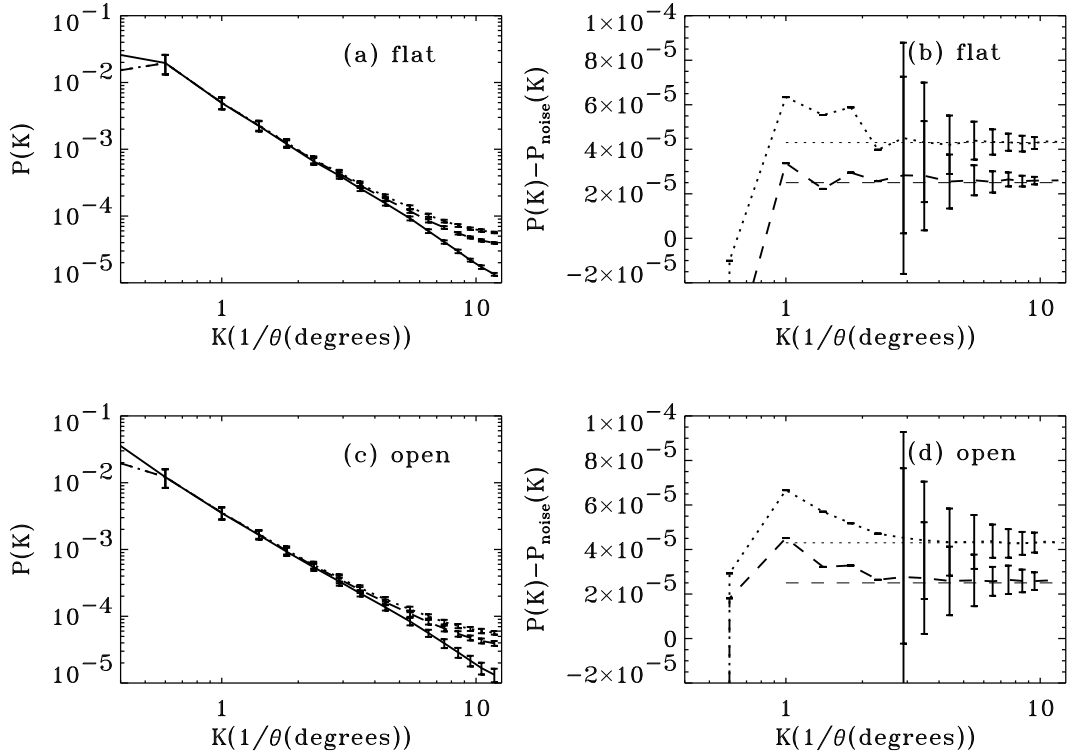,width=15cm}}
\caption{\label{pdek_5x5.ps} Power spectrum analysis of the projected
density field. The upper series of plots are for $\Omega=1$ and the lower series
for $\Omega=0.3$. For all the plots, the solid lines show the power
spectrum of the noise-free mass maps, before any mass reconstruction.
The dotted lines correspond to the power spectrum estimation on the
reconstructed noisy maps with a number density of galaxies of $\bar
n=30~$gal/arcmin$^2$, and the dashed lines for $\bar n=50~$gal/arcmin$^2$.
The left panels show the power spectrum estimates from the reconstructed noisy
maps, compared to the true power spectrum. The 
right panels show the power contribution due to the noise, it is
the difference of power spectra between the
reconstructed noisy maps and the true power spectrum. The thin dashed
and dotted straight lines show the expected value of the noise
contribution in the simple linear noise model described by
Eq.(\ref{power_estim}). It fits remarkably well the true noise level.
}
\end{figure*}

The behavior at scales smaller than our pixel size remains
partly an open issue for two reasons: first,
due to the smaller number of galaxies, the convergence of the reconstruction
process as well as the stability of the noise properties has to be
investigated. Second, at
small scales the gravitational distortion is larger than only a few percent,
and it can go up to infinity on the critical lines. Therefore, estimating the
variance of the galaxies intrinsic ellipticity distribution, arcs and arclets
should be removed. This issue can only be addressed in high resolution
simulations like those performed by Jain et al. (1998). 

\subsection{Power spectrum cosmic variance}

Although the 
estimate of the power spectrum described above is unbiased, the cosmic
variance has also to be explored to come up with an optimal observational
strategy.

In estimating the cosmic variance of the power spectrum,
Gaussian statistics is usually assumed. This hypothesis is tested comparing
the cosmic variance assuming Gaussian statistics to that of the simulated
mass maps and that reconstructed using the two different level of noise
defined before.

In the case of Gaussian statistics
high order moment are related to the second order moments via the
following relations:
\begin{eqnarray}
\langle \tilde\kappa(\kg_1)...\tilde\kappa(\kg_{2p+1})\rangle&=&0, \cr
\langle \tilde\kappa(\kg_1)...\tilde\kappa(\kg_{2p})\rangle&=&
{\displaystyle \sum_{\rm perm}
\prod _{i=1}^p} \langle \tilde\kappa(\kg_{2i-1})\tilde\kappa(\kg_{2i}) \rangle,
\end{eqnarray}
which means that physically, the frequencies of a Gaussian field
are not coupled, and that the
$2p$ moment at a given frequency is only determined by the power at that 
frequency. 

\begin{figure*}
\centerline{
\psfig{figure=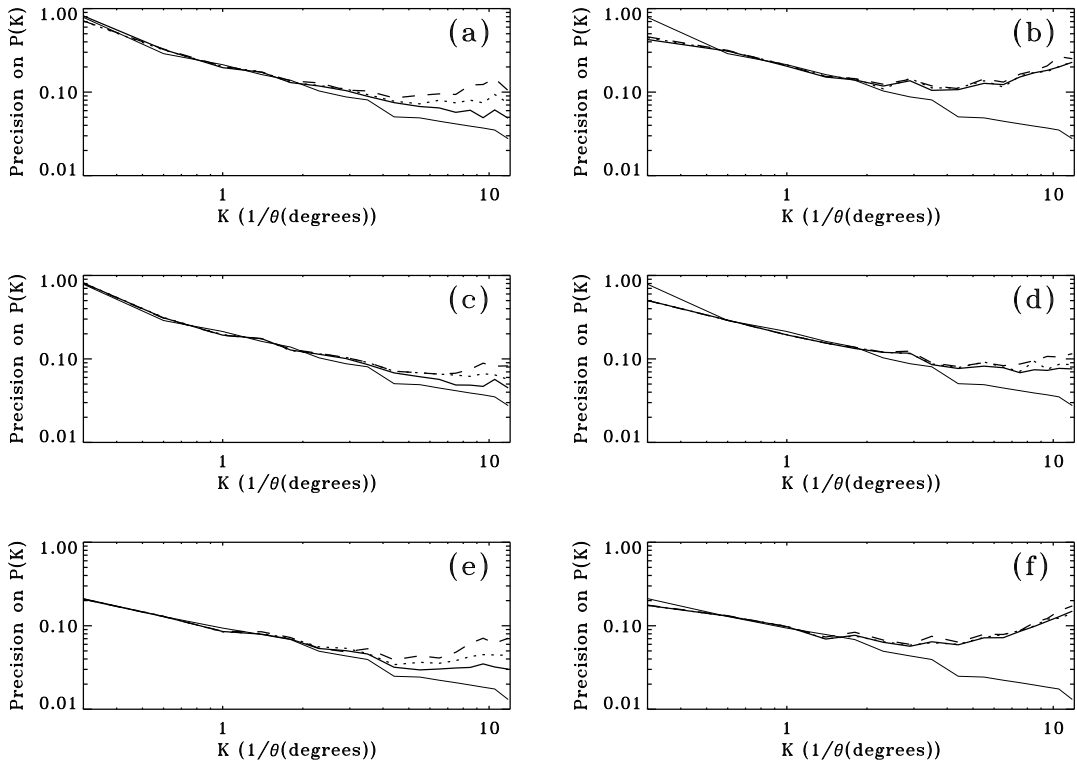,width=15cm}}
\caption{\label{var_pdek.ps} Cosmic variance for flat (left panels)
and open (right panels) models. The thin solid lines are the cosmic
variance expected for a Gaussian density field given by
Eq. (\ref{gauss_cosvar}). The thick solid lines show the true (measured on
the simulations) cosmic variance. The dotted and dashed lines show the
cosmic variance on the reconstructed mass maps, respectively with $\bar
n=50~$gal/arcmin$^2$ and $\bar n=30~$gal/arcmin$^2$. 
}
\end{figure*}

We consider a compact survey of size $\Theta$, for which the
number of modes available at a frequency is maximum. Thus, following
Feldman et al. (1994) and Kaiser (1998), the cosmic variance 
$\sigma_{P_\kappa}^2(\rmk)$ of $P_\kappa(\rmk)$ is given by the 
square of the measured signal
$\langle \tilde\kappa^2(\rmk)\rangle^2$ (which depends on
$\bar\sigma_\epsilon$), divided by
the number of independent modes used to determine it,
$\Delta N(\rmk)$ in the k-annulus $(\rmk,\rmk+\Delta\rmk)$. Simple modes count
gives
\begin{equation}
\Delta N(\rmk)=\pi {\rmk\ \Delta\rmk\over \rmk_0^2},
\label{Nmodes}
\end{equation}
where ${\rm k}_0=2\pi/\Theta$ is the fundamental frequency, thus
\begin{equation}
\sigma_{P_\kappa}(\rmk)={2\sqrt{\pi} \langle \tilde\kappa^2(\rmk)\rangle\over \Theta \sqrt{\rmk\ \Delta\rmk}}.
\label{gauss_cosvar}
\end{equation}
%FB
Note that in this hypothesis the cosmic variance in independent
on the amplitude of the fluctuations. This is not the case
when the non-linear couplings are taken into account
as it can be seen in Fig. \ref{var_pdek.ps}.
It shows the cosmic variance for flat ((a), (c), (e))
and open (((b), (d), (f)) cosmological models. (a) and (b) correspond to
a $5\times 5$ degree survey with $z_s=1$, (c) and (d) with $z_s=1.5$, and
(e) and (f) a $10\times 10$ degree survey with $z_s=1$. On each plot, the
thin solid line shows the Gaussian cosmic variance, the thick solid line
shows the true cosmic variance without noise, 
the dotted line the noisy maps with
$\bar n=50~$gal/arcmin$^2$ and the dashed line with $\bar n=30~$gal/arcmin$^2$.
The vertical axis gives the error on the power spectrum measured at a
given scale. 

The departure from Gaussianity appears for scales below $10'$, the effect
is however more important in the open case model for which non linearities
are stronger. Open models ((b), (d), (f)) give almost the same features
as for the flat models (((a), (c), (e)), although the cosmic variance is
smaller. This is clearly a consequence of a higher
power spectrum signal at low scales for these models, which is visible
when comparing Fig. \ref{pdek_5x5.ps} (a) (flat) and (c) (open). Thus,
as expected,
the intrinsic shape of the power spectrum affects the cosmic variance (Kaiser
1998).

Going deep in redshift (by comparing (a) and (c), or (b) and (d) for the
open case) clearly improves the cosmic variance at small scale, since the
gravitational distortion is stronger.
However this stronger distortion does not improve the large scale power
estimation because the cosmic variance at these scales only depends on the 
whole volume survey.

If shorter wave vectors are observed, (for a $10\times 10$ degree survey),
which is visualized on (e) and (f), a gain of $2$ is reached at all scales,
as a direct consequence of global increase of the number of modes in
Eq.(\ref{Nmodes}). For small scales, from the point of view 
 of the statistics it is in fact equivalent to observe
deep over a small area than to observe shallower over a large area (which
reflects the fact that the dashed lines in (e) and (f) ($\bar
n=30~$gal/arcmin$^2$) are almost the same as the dotted line in (a) and (b) 
($\bar n=50~$gal/arcmin$^2$). But on the other hand, the shallow large
survey gives a better estimate of the power at large scales than the deep survey.
Since these two observational strategies require the same total exposure time
it is clear that wide shallow surveys are better than
small deep surveys. As it will be shown in Sect. 4, this
remains true for the high order moments in real space.
Moreover, deep surveys show more and more distant
galaxies for which the redshift distribution is more uncertain.

\section{Moments in real space}

\subsection{Signature of the normalization and non-Gaussian properties}

Weak lensing carries more information than the amplitude and shape
of the dark matter power spectrum.
Fig. \ref{kappa_model.ps} demonstrates that
distortion maps of the same amplitude (and with very similar
power spectrum as can be checked in Fig. 2) can display very different
features. On these maps the variance of the local convergence 
is the same, but the amount of non-linearities is very different.
For low $\Omega$ universes, the same amount of
distortion can be reached only with a rather large value of $\sigma_8$
thus corresponding to a much more evolved dynamics. As a result
the difference between the underdense and the overdense regions
is more pronounced. The 'voids' tend to occupy a much larger area, whereas
the super clusters tend to be sharper. 
These features appear because of the non-linear couplings
contained in the gravitational dynamics. At large scale the use of Perturbation
Theory has proved to be extremely good in predicting the emergence
of such properties. All these
calculations are based on the hypothesis that the initial conditions were
Gaussian, which we will assume as well.

It has already been stressed (BvWM) that
the departure from a Gaussian statistics is described by 
the skewness of the probability distribution function (PDF)
of the local convergence. We will
restrict our analysis on the skewness basically for two reasons:
it is beyond the scope of this paper to explore all possible
indicators of the non-Gaussian properties, and we know that the
approximate dynamics we have adopted reproduces correctly 
the skewness of the local PDF (see Appendix A). In addition the lens-lens
coupling and the Born approximation terms which are known to be small
for the third moment are probably more important for higher orders, and this
requires a complete dedicated work.

Let us summarize the expected results. For a top-hat window
function we expect to have,
\ba
\sigma_{\kappa}&\approx& 0.01\ \sigma_8\ \Omega_0^{0.8}\ 
\left({\theta_0\over 1{\rm deg.}}\right)^{-(n+2)/2}\ z_s^{0.75},\label{var_def}\\
s_3&\equiv& {\mg\kappa^3\md\over \mg\kappa^2\md^2}\approx
40\ \Omega_0^{-0.8}\ z_s^{-1.35},\label{skew_def}
\ea
where $\sigma_{\kappa}$ is the rms value of $\kappa$ at the
scale $\theta_0$, $n$ is the index of the power spectrum,
$\sigma_8$ is the 3D rms density at 8$h^{-1}$Mpc scale  and $z_s$
is the mean redshift of the sources.
The computed skewness $s_3$ is expected to be independent on the
normalization of the power spectrum. It is only weakly dependent on the
shape of the power spectrum as well as 
on the cosmological constant $\Lambda$ (see discussion).
The skewness would then be a very 
robust way\footnote{It is worth reminding that this is only
possible if the redshift of the sources are perfectly known (which
we assume here).} of determining 
the density parameter $\Omega_0$.

Although the moment analysis is generally
performed on the basis of a top-hat filter
there is a priori no reason to limit our investigations
to this filter. In particular SvWJK have proposed
the use of an alternative function, the compensated filter that might
prove more efficient to constrain $\Omega$, with a lower cosmic variance.

\subsection{Top-hat versus compensated filters}

\begin{figure}
\centerline{
\psfig{figure=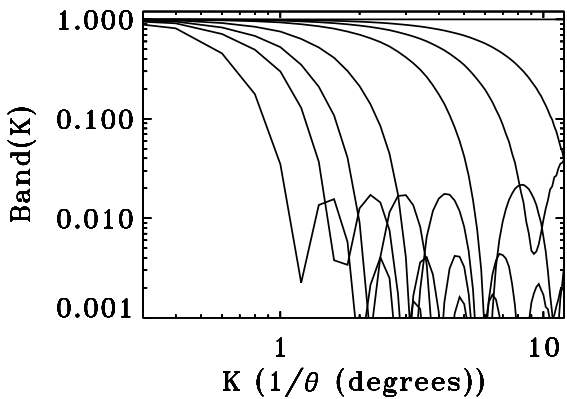,width=7cm}}
\caption{\label{bandshapes.TH.ps} Spectral response for the family of
top-hat filters used in this work. The finite size effects are
visible for the largest smoothing scales. The curves are
not regularly spaced because of pixelisation effects.  
}
\end{figure}

\begin{figure}
\centerline{
\psfig{figure=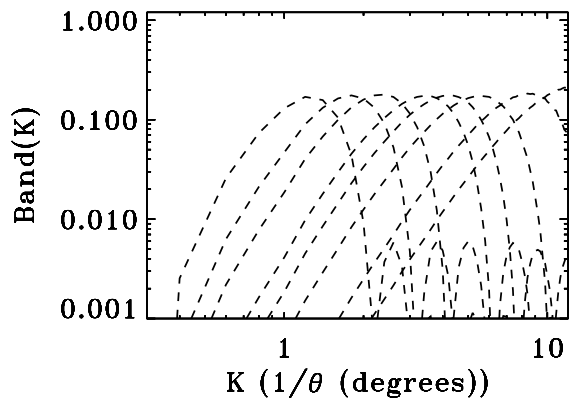,width=7cm}}
\caption{\label{bandshapes.CP.ps} Spectral response for the family of
compensated filters used in this work.
}
\end{figure}

Compensated filters were considered by SvWJK
as a way to measure the convergence directly from the galaxy 
shape. They use the filter $V(\theta)$ of size $\theta_c$ that
defines the quantity $M_{\rm ap}$ as
\begin{equation}
M_{\rm ap}(\thetag)=
\int_0^{\theta_c} \d^2\theta\ V(\theta)\ \gamma_t(\theta),
%\int_0^{\theta_c} \d^2\theta U(\theta)\kappa(\thetag).
\end{equation}
where $\gamma_t$ is the tangential component of the shear field.
$M_{\rm ap}$ is the convergence field filtered by 
$U(\theta)$,
\begin{equation}
M_{\rm ap}(\thetag)=
%\int_0^{\theta_c} \d^2\theta\ V(\theta)\ \gamma_t(\theta),
\int_0^{\theta_c} \d^2\theta U(\theta)\ \kappa(\thetag),
\end{equation}
where $U(\theta)$ has to be the compensated filter and it is related
to the arbitrary filter $V$ through
\be
V(\theta)=-U(\theta)+{2\over \theta^2}\int_0^{\theta}\d\theta'\ \theta'\ U(\theta').
\ee
this latter filter is a compensated filter \footnote{A
compensated filter is a filter with a vanishing mean value. This explains why
$M_{\rm ap}$ can be measured only from the distortion field only since the 
mass-sheet degeneracy is removed by the used type of
filters.} and, for instance, a convenient filter to use is given by,
\begin{equation}
U(\theta)={3\over \pi \theta_c}\left[1-\left({\theta\over \theta_c}\right)^2\right]\left[{1\over 3}
-\left({\theta\over \theta_c}\right)^2\right].
\label{comp_def}
\end{equation}
The spectral responses (defined as the squared amplitude of the Fourier
transforms of the window funstions) 
of the filter's family used in this work are shown in
Fig. \ref{bandshapes.TH.ps} (for the top-hat) and by Fig.
\ref{bandshapes.CP.ps} (for the compensated filter $U(\theta)$).
%FB
Clearly a field smoothed with a top-hat
filter of size $\theta_c$ is sensitive to fluctuations of
size larger than $\theta_c$ that contribute also
to the cosmic variance of the moments. On the other hand, a compensated
filter integrates the fluctuation modes only around the target 
frequency $\theta_c$
and any power at lower or larger scales will affect neither the
signal nor its cosmic variance. SvWJK showed that the de-correlation properties
of compensated filters are by 
far better than for top-hat filters because any power at small wavelengths
between two disconnected fields is highly suppressed (see 
Fig. 8 of their paper).
Thus the cosmic variance should be smaller for a compensated filter
than for a top-hat filter. However this attractive feature comes with
a price: a compensated filter needs to be sampled
by a larger number of galaxies.
In other words, the shot noise has a larger effect
on the aperture mass $M_{\rm ap}$ than on the top-hat filtered mass
at the same scale.
A compromise has to be found, that depends on the functional shape of the 
compensated filter and on the shape of the power spectrum.

\subsection{Moment estimations and shot noise corrections}

Due to the intrinsic ellipticities of the galaxies and the cosmic variance 
(see for instance Szapudi \& Colombi 1996,
Colombi et al. 1998), estimates of the moments of $\kappa$ from the
reconstructed map are biased. We show here that the shot noise can be
accurately calculated and the estimated moments corrected. It is worth noting that   
SvWJK has shown that we can find an unbiased estimator of the moments
of $M_{\rm ap}$, which completely cancel the shot noise correction
problem. It is based on the measurement
of the tangential shear $\gamma_t$. Unfortunately, the measurable
quantity is $g_t$ (the tangential reduced shear)
rather than $\gamma_t$, and unless this is taken into account,
the estimator given in SvWJK is no longer unbiased. In other words, the
shot noise correction problem is shifted to an estimator correction
problem.

It was shown in Section 3.1 that the shot noise leads to a pure white noise
in the reconstructed convergence maps.
The amplitude of this noise can be obtained by measuring the observed
ellipticities of galaxies as described for the power spectrum estimation.
Therefore,  estimates of the variance and the skewness of the convergence in (\ref{var_def})
and (\ref{skew_def}), corrected from the intrinsic ellipticities of the
galaxies, are obtained by simply removing the noise term $\langle
\left(\epsilon^{(s)}\right)^2\rangle$ in the second moment.
Note that the skewness correction only requires the correction of the
variance since the third moment is not affected by the noise.
As the analytical calculation
of the noise term for compensated filters can be rather cumbersome,
it is estimated using Monte-Carlo simulations.

Even after that noise correction, finite sample effects may 
bias the estimations of the second moment and of $s_3$
and may increase significantly the cosmic variance.
This difficulty was partly investigated in BvWM
with the use of perturbation theory. They pointed out that the 
accessible geometrical
averages are expected to be smaller than the true
ensemble averages, and that
a dispersion is expected in the measurements (cosmic variance):
\begin{itemize}
\item the bias that affects the expectation
values was found to be proportional to the variance
at the sample size divided by the one of at the filtering scale;
\item the
scatter was found to be proportional to the rms of $\kappa$ at the
sample size.
\end{itemize}

These estimates were done fully  in perturbation theory, with 
numerous approximations (in particular it was assumed that the sample size
was much bigger than the smoothing scale which is probably
an erroneous approximation for most of the cases considered here).
For accurate investigations of all these effects that take into
account both the Poisson noise and the finite volume effects see 
Szapudi \& Colombi (1996).

\begin{figure*}
\centerline{
\psfig{figure=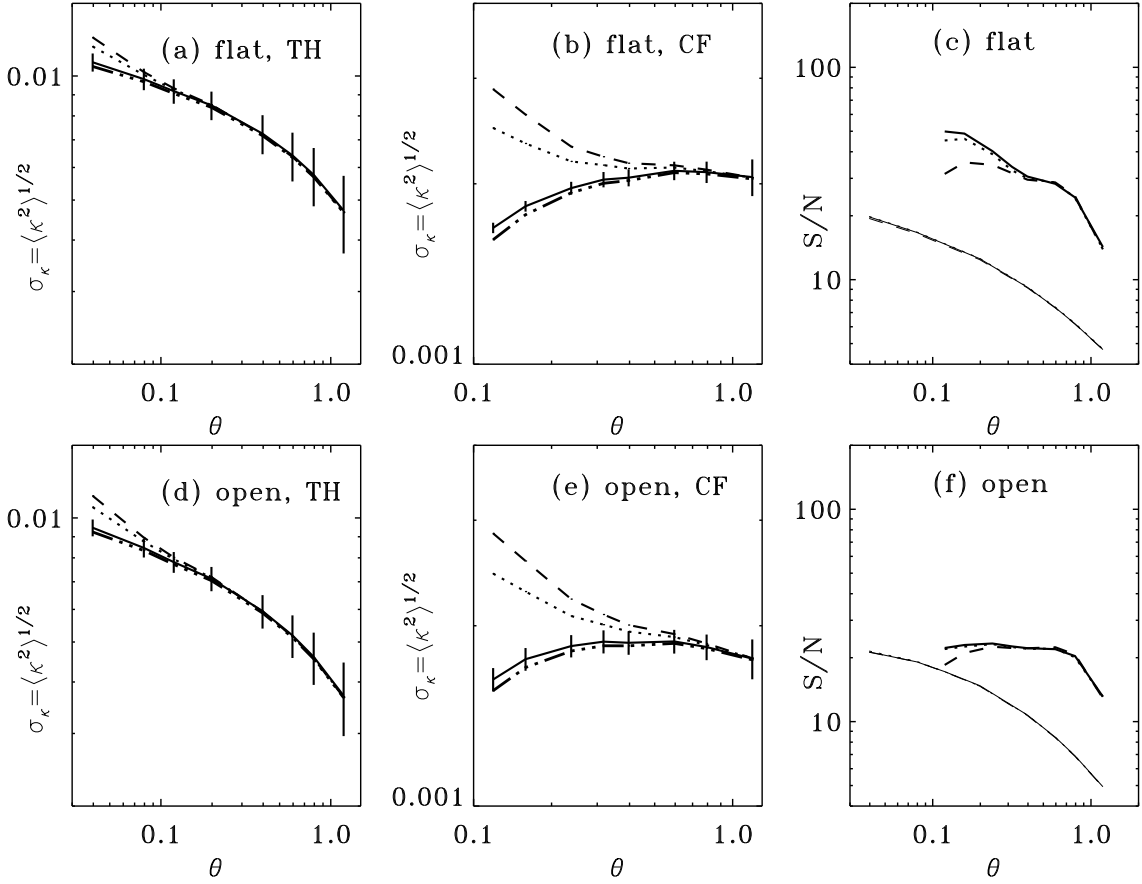,height=12cm}}
\caption{\label{s2_5x5.ps} The measured variance of the convergence, and
the corresponding signal to noise ratio. The upper panels show
the $\Omega=1$ case, while the bottom panels correspond to the
$\Omega=0.3$ case. The left and middle panels are
respectively the measured variance with a top-hat and a compensated
filter. On these plots, the thick solid line is the true variance
measured on the noise-free maps, the dashed line and the dotted line for
the noisy reconstructed mass maps (with respectively $\bar
n=30~$gal/arcmin$^2$ and $\bar n=50~$gal/arcmin$^2$). The dotted-dashed line
is the variance measured from the $n=30~$gal/arcmin$^2$ case and corrected
from the noise. The right panels show the signal to noise ratio of the
variance detection with the top-hat (thin solid line) and
compensated (thick solid line) filters. Dotted and dashed lines have the same
meaning as for (a), (b), (d) and (e) but here the variance has been corrected
from the noise.
}
\end{figure*}

\begin{figure*}
\centerline{
\psfig{figure=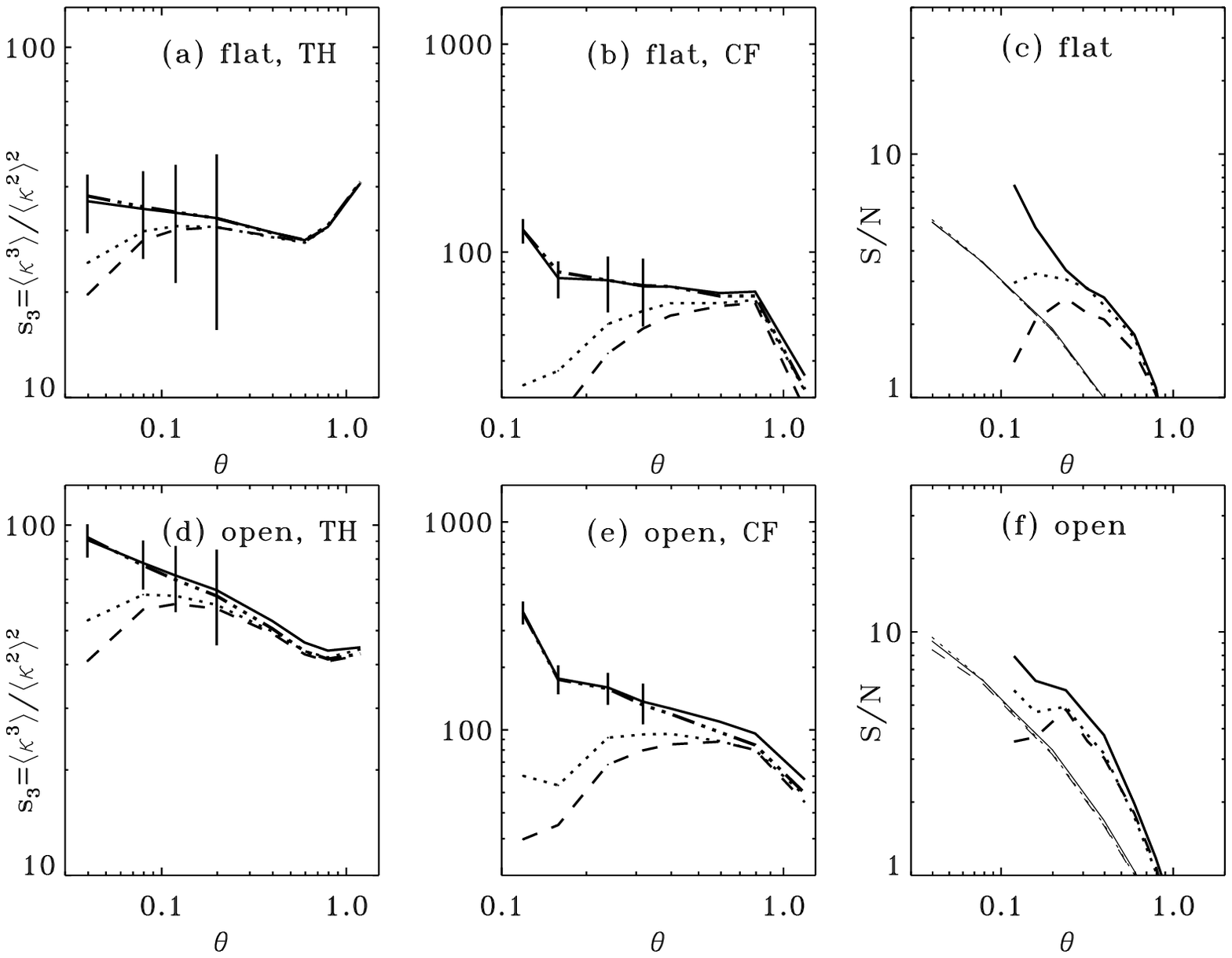,height=12cm}}
\caption{\label{s3_5x5.ps} Same as Fig. \ref{s2_5x5.ps}, but for the
skewness of the convergence, $s_3$.
}
\end{figure*}

\subsection{Results}

We now turn to the measurement of moments in the simulated fields. The
same simulations used for the power spectrum analysis are used here. The
results are given in Figs. \ref{s2_5x5.ps} (variance $\langle\kappa^2\rangle$
in a $5\times 5$ degree field with $z_s=1$) and \ref{s3_5x5.ps} 
(skewness $s_3$ in a $5\times 5$ degree field with $z_s=1$). 
For each of these figures the plots are organized
in the same way: the first raw ((a), (b) and (c) plots) is for the flat model,
the second raw ((d), (e) and (f) plots) for the open model. 
It corresponds to the
first two raws of Table \ref{TSimuls}. The first columns (a) and (d) show
the estimator measured with a top-hat filter, the second columns
(b) and (e) with a compensated filter, and the third columns (c) and (f)
show the signal to noise ratio of these estimators. 
In plots (a), (b), (c) and (d) the solid lines
give the estimators measured in the noise-free $\kappa$ maps, the dotted
lines in the noisy reconstructed maps with
$\bar n=50~$gal/arcmin$^2$ and the dashed lines 
with $\bar n=30~$gal/arcmin$^2$. The dotted-dashed
lines show the estimators measured on the reconstruction with noise
with $\bar n=30~$gal/arcmin$^2$ corrected from the noise.
Since the case $\bar n=50~$gal/arcmin$^2$ gives the same results they are
not plotted. On the signal to noise plots (c) and (f) the thin solid
lines show the results for a top-hat filter and the thick solid lines
for a compensated filter. The results obtained for the noisy maps
with $\bar n=30~$gal/arcmin$^2$ or $\bar n=50~$gal/arcmin$^2$ 
{\it corrected from the noise}
are respectively given by the dashed and dotted lines (either
thin or thick).

\subsubsection{Noise correction}

The noise correction as described in Sect 4.3
gives unbiased results,
as it is expected for superimposed white noise.
This is true even when the correction is two orders
of magnitude higher than the signal (see for example
Figs. \ref{s3_5x5.ps} plots (b)
and (e)). This confirms the
simple properties of the noise in the reconstructed mass maps
already found in Sect. 3.1. 

\subsubsection{The variance}

The variance obtained for a top-hat filter
is slowly decreasing with an increasing scatter, as expected.
For the compensated filter the curves are almost flat
as expected from the shape of the power spectrum.
The noisy maps (dotted and dashed lines in Fig. \ref{s2_5x5.ps}.(a) and (b)
display higher values for the variance. Once it is corrected,
the results are in perfect agreement with the noise-free simulations.
The signal to noise ratio is basically not affected by the shot noise
for a top-hat filter, it shows that going deep does not improve
the measurement precision. The compensated filter reveals much more
sensitive to the shot noise as predicted in section 4.2 since
we can see on plot Fig. \ref{s2_5x5.ps}. (c)
and (f) (thick lines) the bell shape of the signal
to noise ratio, with a significant reduction of the measurement
precision at the smallest available scales.
The open and flat cases show basically no differences. It can be seen
however that the signal to noise ratio for a compensated filter
is sightly lower for the open case. We interpret this effect
as due to the presence of more nonlinear couplings in the maps.
Remarkably, the precision with which the variance can be measured in some
specific $k$ range reaches 5\%.

\subsubsection{The skewness}

The skewness can be accurately
measured at the smallest scales (in Fig. \ref{s3_5x5.ps} only the error
bars for the first four points have been drawn). The skewness is decreasing
with scale in the two cases. Once again, the noise correction applied to the
reconstructed maps allow to recover the skewness with a surprising accuracy.
The signal to noise of the skewness is still not affected by the noise
for a top-hat filter, while for the compensated filter the situation is worse
than for the variance; for instance, at the smallest scale,
the signal to noise is almost one order of magnitude smaller 
on the reconstruction with noise
than on the noise-free maps for the $\Omega=1$ case.
The two models, open and flat, provide us with very different
magnitude for $s_3$.
Their skewness ratio is 3 as expected from 
perturbation theory\footnote{To reduce the cosmic variance, we reject
the pixel values that are above 4 $\sigma$. It slightly 
reduces the  value of $s_3$ for the open case.},
 with a significance of the separation at roughly
$6\sigma$. This is observed in case
of the top-hat as well as the compensated filter and this confirms the
fact that the skewness of the convergence can strongly separate low and
high density universes.

\begin{figure}
\centerline{
\psfig{figure=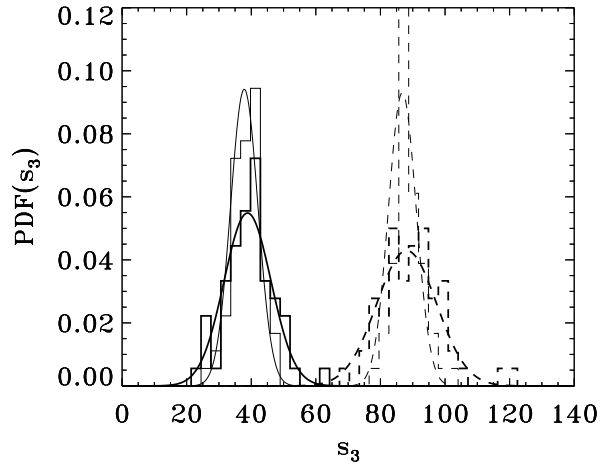,height=7cm}}
\caption{\label{Histo.s3}
Histograms of the values of $s_3$, top-hat filter, 
for $\Omega=1$ (solid lines) and $\Omega=0.3$ (dashed lines)
for a $5\times5$ degree survey (thick lines) and a $10\times10$ degree survey
(thin lines). The angular scale is the pixel size $2.5'$.
}
\end{figure}

To be more precise we present the actual histograms of the measured
skewness in Fig \ref{Histo.s3} which demonstrates clearly that the two
cosmologies can be easily separated.
One can see that the scatter in $s_3$ is roughly the same in the two cases
and that the difference in the relative precision is due to the
differences in the expectation values. This plot also shows that 
the distribution of the measured $s_3$ is quite Gaussian.

\subsection{Comparisons of different observational scenarios}

The results presented previously had been obtained from the full mass
reconstruction of 240 maps \footnote{This corresponds to the two noise levels
for each cosmological model, with 60 mass maps for each case.}
as described in Appendix B.
 Since it was demonstrated in the preceding
section that noise acts as a pure de-correlated white noise in 
the reconstructed
$\kappa$ maps, we pursue our analyses of large series of simulated
fields simply by adding the noise on the initial $\kappa$ maps (especially
for $10\times 10$ degree data sets for which convergence reconstruction
would take typically one and a half hour on DEC PWS-500 computers). 
The subsequent analysis are therefore made with this simplified scheme.

To complement the previous cases, we have built and analyzed the
cosmic variance on 60 maps for each of the following models: open
($\Omega=0.3$) and flat ($\Omega=1$) cosmologies, a survey size of $5\times 5$
degrees for $z_s=1.5$, a survey size of $10\times 10$ degrees
for $z_s=1$, with the power spectrum of Eq. (\ref{BGpower}) and 
a survey size of $5\times 5$ for $z_s=1$ with a CDM spectrum.

\begin{figure*}
\centerline{
\psfig{figure=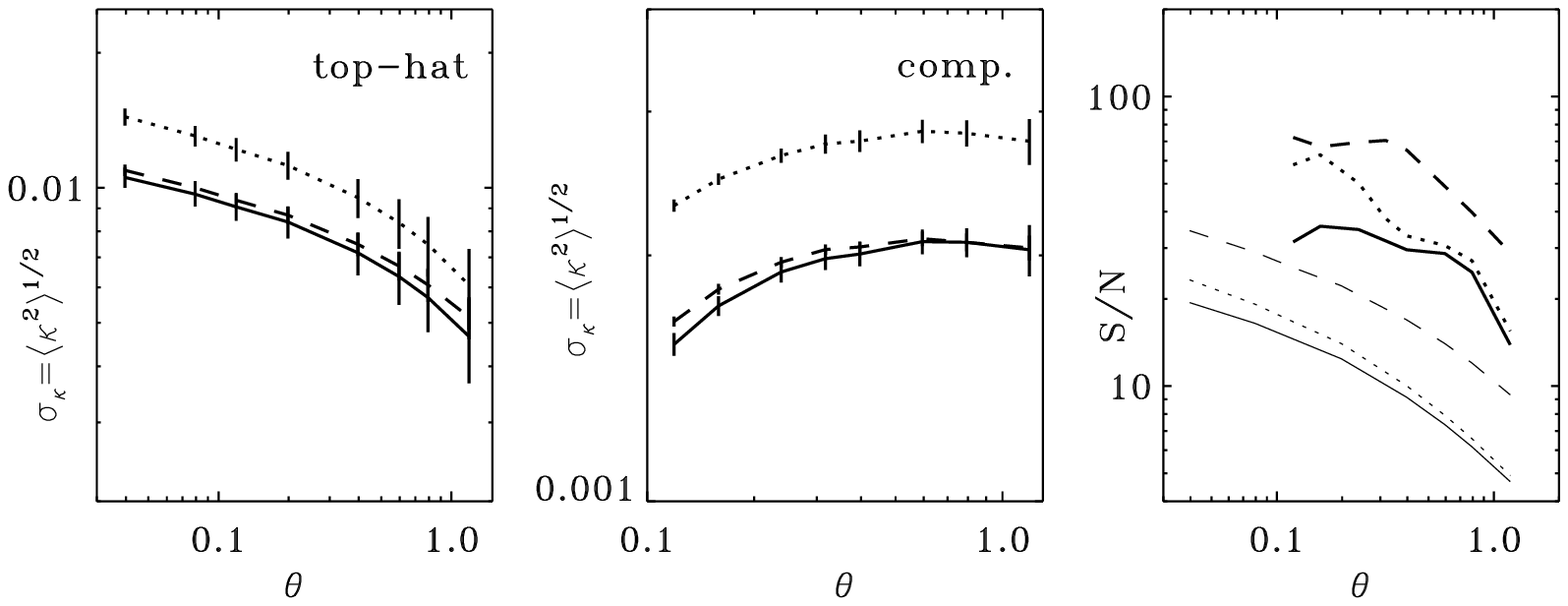,height=12cm}}
\vspace{-6cm}
\centerline{
\psfig{figure=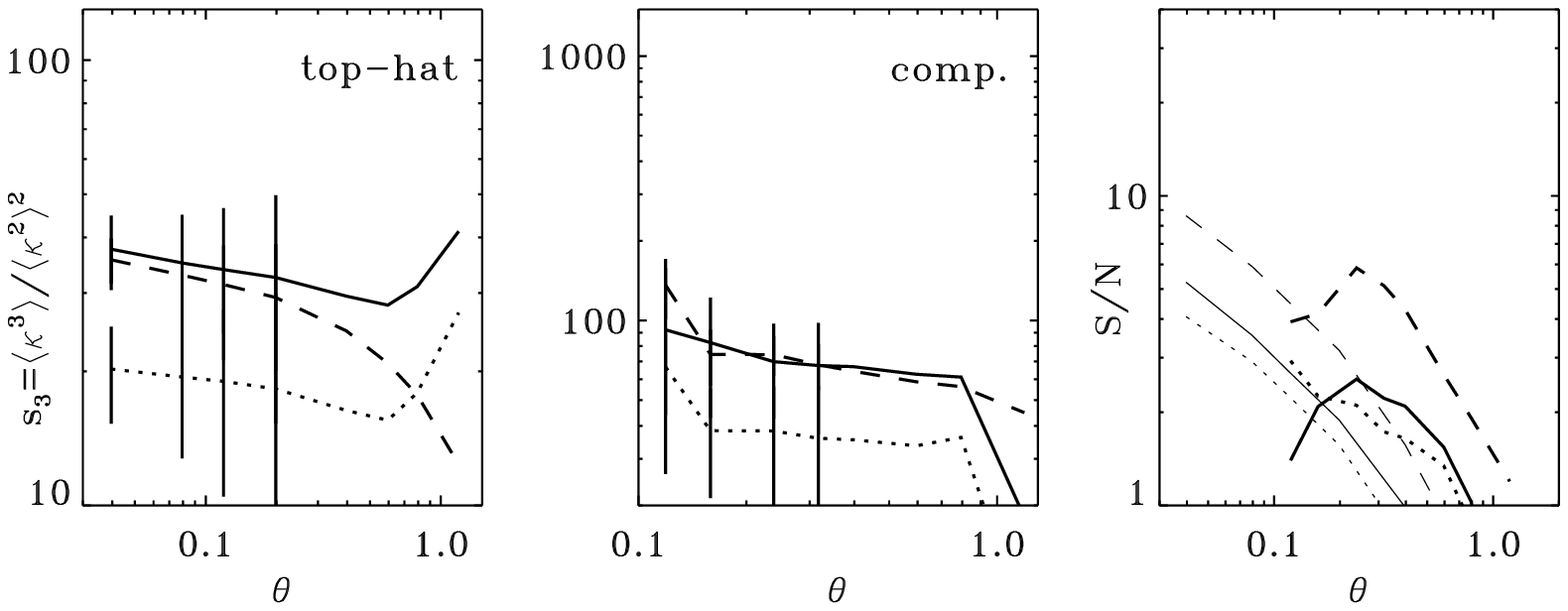,height=12cm}}
\vspace{-6cm}
\caption{\label{s23_comp.ps} For a flat cosmological model, comparison
of different observational strategies
between a survey of size $10\times 10$ degrees (dashed lines) a $5\times 5$
degrees survey for sources at mean redshift 1.5 (dotted lines) and a
$5\times 5$ degrees survey for sources at mean redshift 1 (solid lines).
Left panels are for the top-hat filter and middle panels for the
compensated filter. The thicker lines in the right panels hold for the
compensated filter.
}
\end{figure*}

\subsubsection{Effect of the survey size}

By increasing the total area of a factor $4$ we increase the signal to noise
on the variance and the skewness with a top-hat filter by 
almost a factor $1.7$. This can be seen in Fig. \ref{Histo.s3}
when comparing the $5\times5$ degree case with the $10\times10$ case.
With a compensated filter, 
the signal to noise ratios of the variance and of the 
skewness are increased by exactly a factor 2, thus improving more
rapidly with the sample scale than the top-hat window function. This
is expected from the de-correlation properties of those filters.
It makes the compensated filter actually more attractive for
such a large survey.

\subsubsection{Effect of the source redshift}

Fig. \ref{s23_comp.ps} shows the effect of a change in
the mean source redshift.
For the galaxies that are further away, 
the variance of $\kappa$ is larger since the gravitational distortion
is stronger, conversely the skewness is smaller since the
accumulated material along the line of sight creates a field that is
 more and more Gaussian.
The surprising result is that the signal to noise of these quantities does not
depends strongly on the redshift of the sources. This means that it will be
a waste of time to observe at high redshift, while it will basically not
improve the precision of the measurement. Things are slightly improved
for the compensated filter, but fundamentally, the results with the top-hat
filter show that we do not learn more by increasing the redshift. Note that
there is no improvement due to the increasing galaxy number density if the
survey size is unchanged (see section 4.3). In
addition, high redshift surveys may create new problems such as uncertainties
due to the Born approximation, the lens-lens coupling or the recently
investigated source clustering effect (Bernardeau 1998).

\begin{figure*}
\centerline{
\psfig{figure=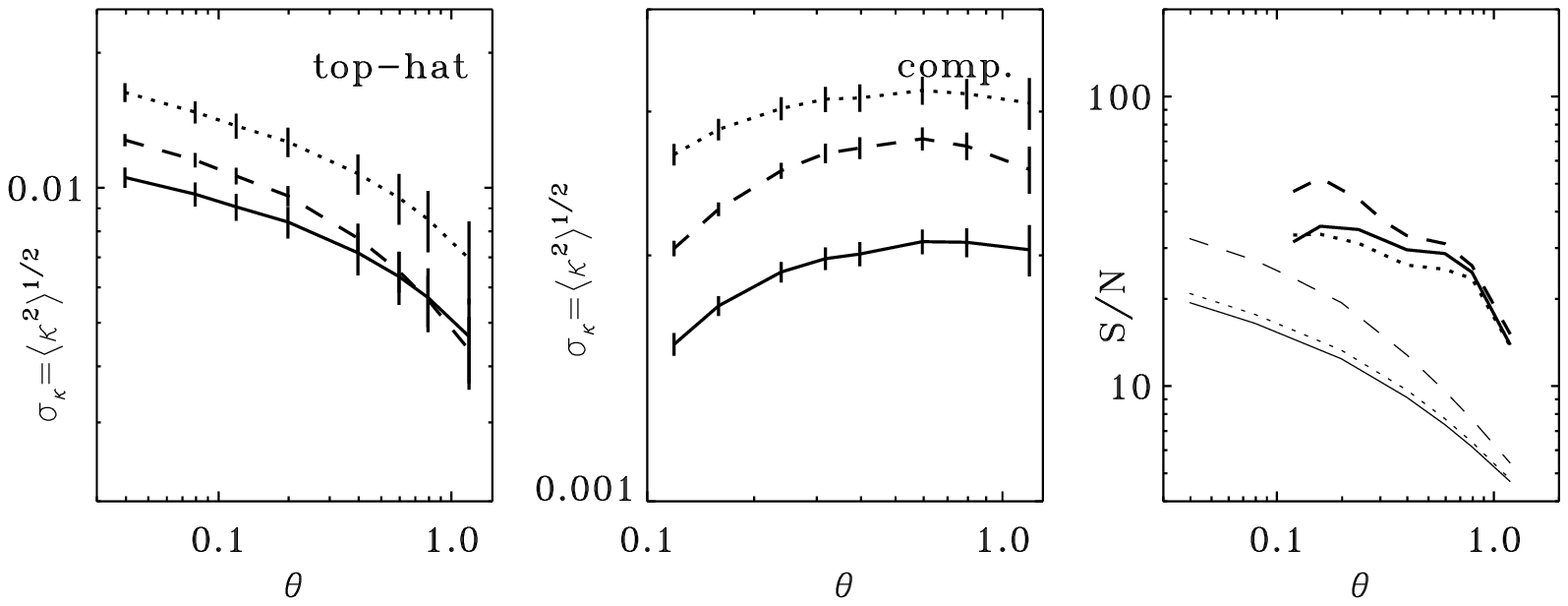,height=12cm}}
\vspace{-6cm}
\centerline{
\psfig{figure=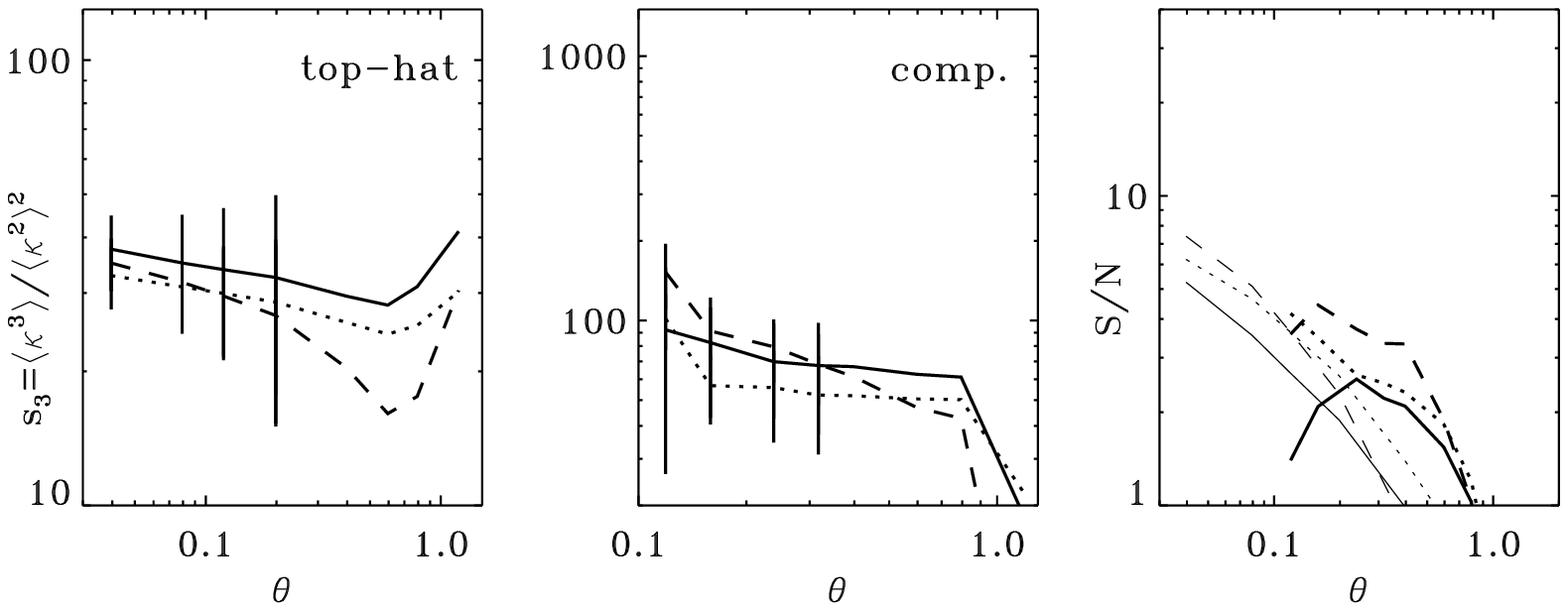,height=12cm}}
\vspace{-6cm}
\caption{\label{s23_power.ps} Comparison of different choices of
power spectra for a $5\times 5$ degree survey and a flat cosmological
model. The dashed lines
correspond to an $\Omega=1$ CDM spectrum with $\sigma_8=0.6$ and $z_s=1$,
dotted lines to a BG spectrum with $\sigma_8=1$ and $z_s=1$, and the solid
lines to a BG spectrum with $\sigma_8=0.6$ and $z_s=1$.
}
\end{figure*}

\subsection{BG versus CDM power spectrum}

In order to test the robustness of the skewness as an estimator of $\Omega$
independent on the power spectrum we re-ran our simulations
for a standard CDM power spectrum. Fig. \ref{s23_power.ps} shows
the comparison between the  CDM model
(dotted line) and BG power spectra, which clearly shows that CDM
contains more structures at small scale by looking at the variance plots (a),
(b). As predicted in BvWM, 
the skewness of the convergence if almost unaffected by the change of power
spectrum (see Fig. \ref{s23_power.ps} (d), (e)), but there is a small
improvement in the signal to noise for the flat model (because of the
larger power at small scale for CDM). On the other hand, the variance
is strongly affected and in particular there is a significant  decrease of
power at large scale compare to
BG spectrum. Note that the compensated filter yields a more accurate
representation of the underlying power spectrum
than the top-hat filter (because it is a pass band filter), and leaves the
angular dependence of the variance unchanged compared to the spectrum
Eq.(\ref{BGpower}) excepts at large scales.

The skewness is thus a robust estimator of $\Omega$ fairly insensitive to the
power spectrum.

\begin{figure}
\centerline{
\psfig{figure=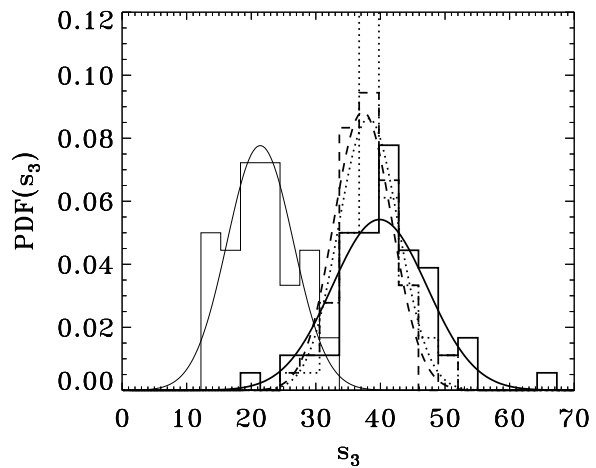,height=7cm}}
\caption{\label{Histo.flats.s3}
Histograms of the values of $s_3$ for $5\times5$ degree survey, 
top-hat filter, 
for $\Omega=1$ for a BG spectrum with $\sigma_8=0.6$ and $z_s=1$
(thick solid line), $\sigma_8=0.6$ and $z_s=1.5$ (thin solid line), 
$\sigma_8=1.0$ and $z_s=1$ (dotted line) and a CDM spectrum with
$\sigma_8=0.6$ and $z_s=1$ (dashed line). The angular scale is the
pixel size, $2.5'$.
}
\end{figure}

\subsection{Effect of the normalization}

The skewness 
$s_3$ is found to be independent on the normalization, as expected from the
Perturbation Theory. The signal to noise ratio however is increased by 
about 40\% in the case of high normalization $\sigma_8=1$. These
results are summarized in Fig. \ref{Histo.flats.s3} that shows the
histograms for various cosmological cases. It demonstrates that the
skewness is clearly independent on the shape
and normalization of the power spectrum. However there is a strong
dependence on the mean redshift of the sources. If the signal to
noise ratio for $s_3$ depends on the cosmology it is 
independent on the mean source redshift. This again
is favoring rather shallow surveys.

\subsection{Beyond the skewness to measure $\Omega$?}

The skewness of the PDF of the local convergence does not entirely
characterize the PDF itself. Thus it is natural to measure higher order 
moments of the convergence to probe the cosmology. 

It is clear that the skewness
breaks the degeneracy between the power spectrum and the cosmological
parameters, and is completely insensitive to the normalization.
Bernardeau (1995) already noticed that in the case of cosmic density field
or cosmic velocity field, the ratio $s_4/s_3^2$ calculated from the perturbation
theory with a top-hat filter is almost a constant, and
independent on the underlying cosmological model. This work was recently
extended to the lensing case (Bernardeau, 1998) where he found $s_4(\kappa)
/s_3(\kappa)^2\simeq 2$. If all systematics of the gravitational lensing
measurement can be controlled, search for such a {\it magic} number in our
Universe would be a strong indication of validity of the paradigm of the
gravitational instability scenario started from Gaussian initial conditions.
On the other hand $s_4$ may be a new
way to measure the density parameter (but not totally
independent on the skewness).
In our maps we find that the noise correction
still works for the kurtosis, and that a compensated filter is more
efficient than the top-hat filter (at least for noise-free data). 
In addition the kurtosis appears
to be a fairly good discriminant for $\Omega$. Unfortunately the
signal to noise ratio remains lower than for the skewness, which make
it more difficult to measure.
Moreover, it is more sensitive to the usual
lensing approximations (Born approximation and lens-lens coupling terms)
as well as source clustering (Bernardeau 1998). The error bars of the kurtosis
found in our simulations
are so large with a $5\times 5$ degree survey that it is impossible to
measure it at scales larger than a few arcminutes.
It turns out that the ratio $s_4(\kappa)/s_3(\kappa)^2$ is $\sim 1.4$ for the
flat case and $\sim 1.8$ for the open case for a top-hat filter, while it
is $\sim 5.6$ for the flat case and $\sim3.4$ for the open case for a
compensated filter. It is not the scope of this work to compare further
the differences between the two filters, but we want to point out that
filtering may be an interesting way to change the dependence of an estimator
versus the cosmology. This should be studied in order to search for
optimal measurements of higher order moments. More generally, the
whole shape of the PDF could probably be used with more efficiency
than the skewness alone. In a regime of small departure from a
Gaussian distribution it can for instance be fruitful 
to describe the shape of the PDF with an Edgeworth expansion that
takes into account the first few moments (Juszkiewicz et al. 1995, 
Bernardeau \& Kofman 1995). %FB
For instance, from the Edgeworth expansion it is easy to show
that the fraction of values of the convergence that is above 
the average value, $P(\kappa > \mg<\kappa\md>)$, is
\begin{equation}
P(\kappa > \mg<\kappa\md>)\approx {1\over 2}\ \left(
1-{s_3\,\sigma_{\kappa}\over 6\sqrt{2\pi}}\right).
\end{equation}
The results obtained from this formulae are in good agreement
with those obtained from the direct measurements, however with
a slightly larger cosmic variance.

Finally, the non-Gaussian features can also be characterized with
topological indicators (which have been shown to be fruitful for the
analysis if CMB data already, see for instance Winitzki \&  Kosowsky 1998,
Schmalzing \& Gorski 1997). What cosmic variance could be derived
from the joint use of topological quantities or/and information
on the shape of the PDF is left for further investigations.

\section{Discussion}

In this paper, we have focussed our investigations on scales larger than
$2.5'$ thus allowing ourself the use of a simplified dynamics.
This  is a complementary approach to the ongoing
investigations by Jain et al. (1998) who analyzed the high order moments
and the power spectrum of the calculated from ray tracing in high
resolution simulations. Although they only analyze the statistical properties
of the convergence without noise, their approach completes our own
towards the smaller angular scales.
 They have shown in particular that the skewness of the convergence is
significantly higher at small scales ($0.1$ arcmin) than
the theoretical expectations of perturbation
theory due to highly non-linear structures, (this was already mentioned by
Gazta\~naga \& Bernardeau 1998 and whether this behavior can still
be described by means of perturbation theory with the
help of loop correction terms is still an open question). 
Unfortunately, even at these small scales, they are
not able to analyze the cosmic variance because of the small number
of realizations. The use of 
high resolution simulations probably prevents a detailed analysis
of this quantity.

The preceding sections provide quantitative 
estimates of the capability of mass reconstructions from weak lensing
measurements at large scale  to probe the large
scale structures, as well as the cosmological parameters. We have shown how
the projected mass distribution can be reconstructed accurately from
the observed shape of the galaxies. Two complementary analysis have
been examined, the power spectrum and the non-Gaussian features
through the high order moments.

For the power spectrum estimation, the best sampling strategy (i.e. the
question of sparse or compact surveys) is not
discussed in this paper (see Kaiser 1998 for a discussion),
but our results show that in order to probe the smallest scales of mass
fluctuations a deep (but narrow) survey 
is required that  diminishes the cosmic
variance caused by the shot noise.
Once this is done, the survey can be extended in a more shallower
manner to probe the power spectrum at scale where 
the cosmic variance caused by shot noise becomes unimportant.
At this stage, the question
of a sparse or compact survey is a matter of choice, depending
on scientific interests.

\begin{table}
\caption{Relative error on the measured variance and skewness of the
convergence in
different observational contexts (in  tables $n_1=30~$gal/arcmin$^2$ and 
$n_2=50~$gal/arcmin$^2$) for a top-hat filter. The numbers correspond
to the smallest error within the range of scales considered in this
work.
}
\label{T1}
\begin{center}
\begin{tabular}{|c|c|c|c|c|c|}
\hline
\multicolumn{2}{|c|}{Observational}&\multicolumn{2}{|c|}{$\Omega=1$}&\multicolumn{2}{|c|}{$\Omega=0.3$}\\
\cline{3-6}
\multicolumn{2}{|c|}{constraints}&$\delta \sigma_{\kappa}/\sigma_{\kappa}$&$\delta S_3/S_3$&$\delta \sigma_{\kappa}/\sigma_{\kappa}$&$\delta S_3/S_3$\\
\hline
$5\times 5$ deg&$n_1$&0.051&0.190&0.047&0.118\\
\cline{2-6}
$z_s=1$ &$n_2$&0.050&0.186&0.047&0.105\\
\hline
$5\times 5$ deg&$n_1$&0.043&0.246&0.038&0.127\\
\cline{2-6}
$z_s=1.5$ &$n_2$&0.043&0.223&0.038&0.123\\
\hline
$10\times 10$ deg&$n_1$&0.029&0.116&0.024&0.060\\
\cline{2-6}
$z_s=1$ &$n_2$&0.029&0.110&0.023&0.054\\
\hline
\end{tabular}
\end{center}
\end{table}

\begin{table}
\caption{Same as Table \ref{T1} for a compensated filter
}
\label{T2}
\begin{center}
\begin{tabular}{|c|c|c|c|c|c|}
\hline
\multicolumn{2}{|c|}{Observational}&\multicolumn{2}{|c|}{$\Omega=1$}&\multicolumn{2}{|c|}{$\Omega=0.3$}\\
\cline{3-6}
\multicolumn{2}{|c|}{constraints}&$\delta \sigma_{\kappa}/\sigma_{\kappa}$&$\delta S_3/S_3$&$\delta \sigma_{\kappa}/\sigma_{\kappa}$&$\delta S_3/S_3$\\
\hline
$5\times 5$ deg&$n_1$&0.028&0.391&0.044&0.209\\
\cline{2-6}
$z_s=1$ &$n_2$&0.021&0.310&0.043&0.174\\
\hline
$5\times 5$ deg&$n_1$&0.016&0.341&0.030&0.193\\
\cline{2-6}
$z_s=1.5$ &$n_2$&0.016&0.310&0.029&0.172\\
\hline
$10\times 10$ deg&$n_1$&0.014&0.171&0.020&0.110\\
\cline{2-6}
$z_s=1$ &$n_2$&0.014&0.140&0.020&0.074\\
\hline
\end{tabular}
\end{center}
\end{table}

Concerning the moments measurement, a summary is given in Table
\ref{T1} (for a top-hat filter) and \ref{T2} (for a compensated
filter) which shows the smallest accessible error on the  measurement
of the variance and the skewness of the convergence for different
observational contexts. This demonstrates that weak lensing
measurements can reach a few percent precision on $\Omega$ for a
reasonable survey size.  We pointed out that the
results are not much deteriorated for a number density of galaxies of
$30$ gal/arcmin$^2$ compared to $50$ gal/arcmin$^2$, whereas the
realization of the survey in the latter case requires a factor 3 more
observing time.  It suggests that large shallow surveys would be more
adequate since it efficiently reduces the cosmic variance. Such a
strategy would be much more comfortable with respect to some
systematics like the redshift distribution of the sources (which can
be determined easily if the sources are closer), the source
clustering effects  that are limited if the source distribution is
narrow (Bernardeau 1998), and possibly the morphological evolution of 
distant galaxies, in particular if most distant galaxies are composed 
of many  merging substructures.

Note that there is still room for potential improvement of the signal
to noise ratio we are obtaining.  Indeed, since we are limited by
construction to simulations of scales larger than $2.5'$ we do not
know whether the cosmic variance  can be reduced by observing the
moments at smaller scales, as it is suggested by the results obtained
with a top-hat filter (for which the signal to noise curves never bend
down at small scales).  The optimal size for the measurement of the
variance and the skewness might in fact correspond to the arcmin
scale. However, there are several issues that are generally believed
to be irrelevant for weak lensing at large scales, but are probably
potential difficulties at small scales:
\begin{itemize}
\item Our work implicitly assumes a constant power spectrum
below the pixel size.
A mass reconstruction from real data should include small scale
features such as cluster lensing, and the propagation of the noise from this
peaks of signal is not known.  Moreover that ability of the $\chi^2$
method to match the noise properties correctly at small scales
should be reinvestigated.
\item Born approximation and lens-lens coupling terms are stronger at small
scales (see SvWJK). A quantification of this effect on simulations would be necessary
to decide at which
scale the measurement of moments is optimal.
\item Source clustering (investigated at large scales by means
of perturbation theory by Bernardeau 1998) could also have a significant
impact at small scale.
\end{itemize}

The possible systematics caused by all these effects have to be
investigated in order to estimate the precision of weak lensing surveys.

Moreover the moments are not
necessary the {\it best} means to distinguish
between different cosmological models (not to mention the fact that 
the results are anyway sensitive to the choice of the filter). 
For example, it seems that by comparing open and flat
universes on Fig. \ref{kappa_model.ps} topological tools should be
as well a strong discriminant of cosmological models. 
The statistical instruments are extremely diverse. Among the possible tools
the results of multi-scale filtering, as provided by wavelet transforms,
is likely to be a good candidate. 

\section{Conclusion}

We have investigated in detail how weak lensing observations could 
be used to measure the projected power spectrum and to
discriminate among different cosmological models.
In order to analyze the cosmic variance of the results
(which requires a large
number of realizations), we used a two dimensional, second order Lagrangian
dynamics to generate the convergence fields. In each of the
observational situation and cosmological model, 
60 different maps has been generated. Cosmological
models include: flat model ($\Omega=1$), open model ($\Omega=0.3$), Baugh \&
Gazta\~naga initial power spectrum, standard CDM spectrum. Observational
contexts include: high noise ($30~$gal/arcmin$^2$) and low noise
($50~$gal/arcmin$^2$) level, "small" survey size ($5\times 5$ degrees) and
large survey size ($10\times 10$ degrees), two different mean redshift
of the sources ($z_s=1$ and $z_s=1.5$), and the use of top-hat and compensated
filters. From these convergence maps, a shear map is calculated, on which
the noise is introduced according the selected observational context, 
the mass map is finally reconstructed (using a $\chi^2$ method) and analyzed.

Our results are:
\begin{itemize}
\item At scales larger than $2.5'$ the $\chi^2$ reconstruction method
is a very stable process which
does not produce any boundary effects, spurious signal, and which leaves
the noise properties unchanged (the noise does not propagates and remains close
to the theoretical prediction using the linearized lens equation). This
permits to work with the convergence (which is
the physical field of interest) instead of the shear, with no loss of
information.
\item The shot noise contribution can be removed simply by measuring
the {\it observed} ellipticities of the galaxies, and this leads to unbiased
estimates of the power spectrum and the moments of the convergence
(at least up to the kurtosis).
\item  The precision obtained on the normalization can be as low as 2\% 
with survey of $10\times 10$, and 5\% for $\Omega$ (see tables 1 and 2)
for the most favorable cases (i.e. low $\Omega$).
\item The larger shallower surveys are more promising to recover the cosmological
quantities. In addition to that,
the redshift distribution of the brighter galaxies is known better and their
light distribution is more regular, making ellipticity measurement from their weighted 
 second moment easier and more relevant.
\item The compensated filter yields by far smaller cosmic variance than the
top-hat filter for the variance of the convergence, while it gives
unsatisfactory results
for the skewness. However the cosmic variance decreases more rapidly with an
increasing survey size for a compensated filter.
\end{itemize}

The MEGACAM project offers the possibility to perform these observations.
Indeed the large field of the CCD device (1 square degree) and the high image
quality at CFHT (extended to the edges of the field with the future field
corrector) provides the ideal instrument to perform
such a scientific program.

%As an example, one night should be enough
%to observe a $2\times 2$ degrees area, which leads to a 95\% precision
%on the power spectrum normalization (????) and to ...\% on the skewness. This
%will in addition gives the first indication of the large scale mass
%distribution, as we can see on Fig. \ref{kappa_model.ps}.

\begin{figure}
\centerline{
\psfig{figure=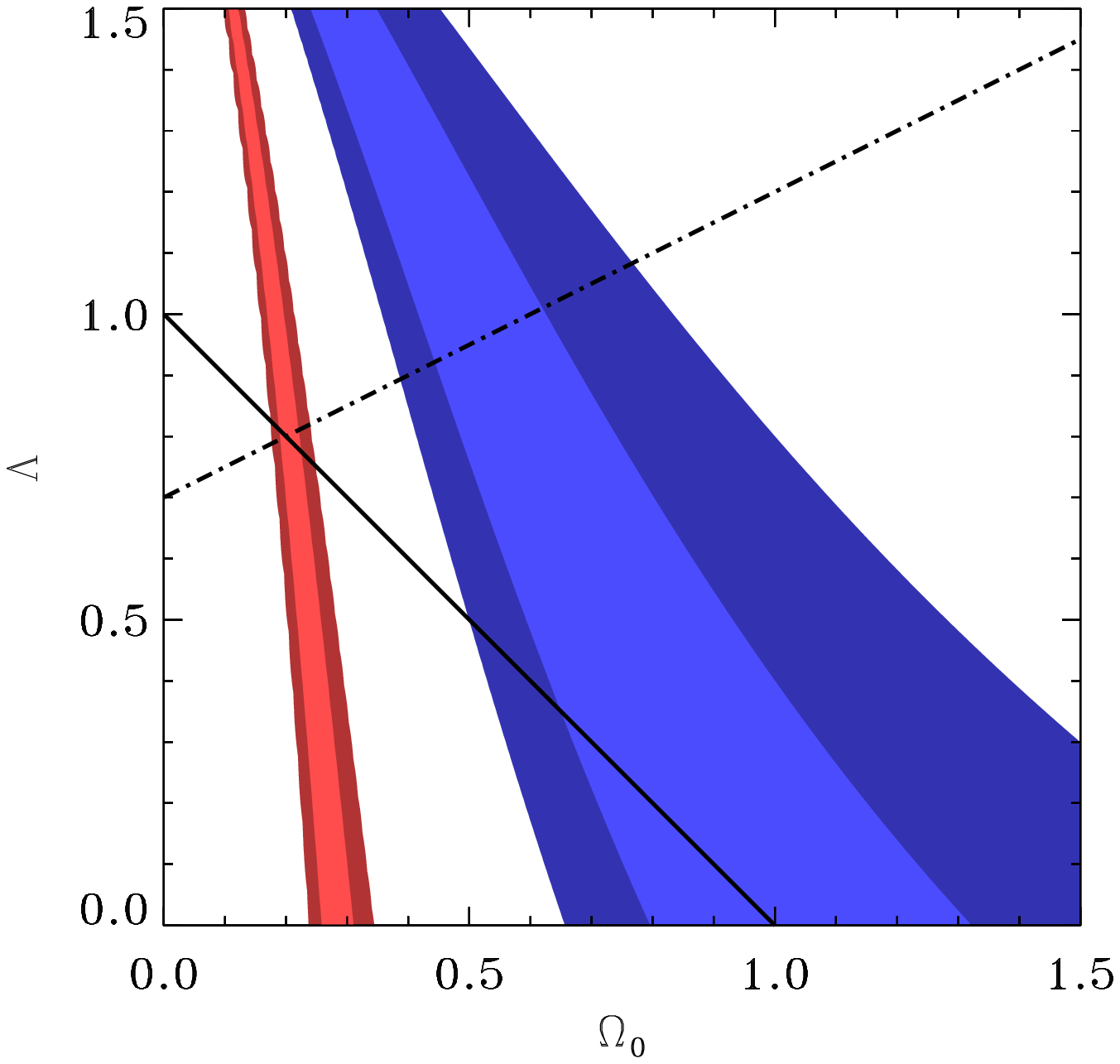,width=8cm}}
\centerline{
\psfig{figure=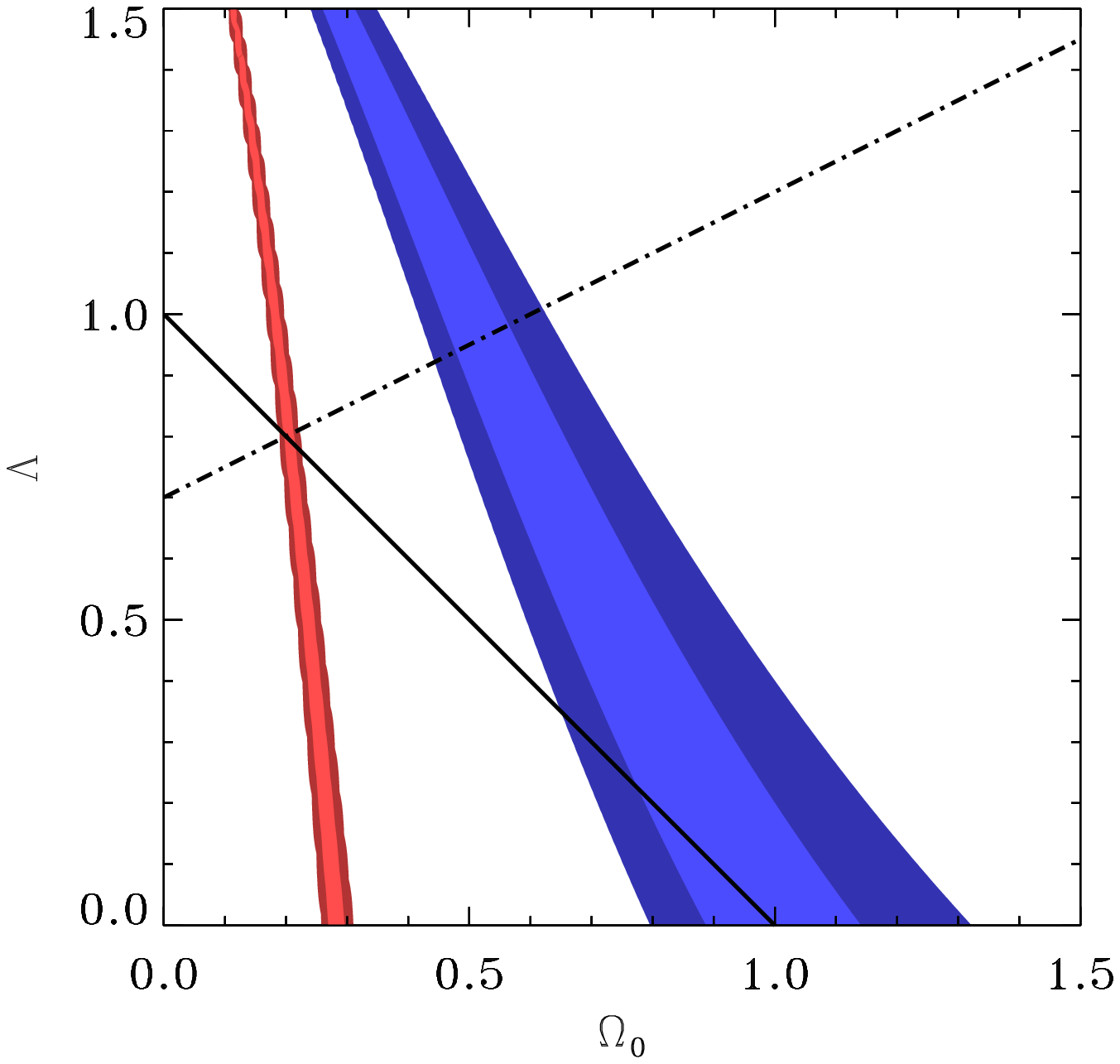,width=8cm}}
\caption{\label{OmLambda.ps} Constraints that can be brought 
by weak lensing survey in an $\Omega_0-\lambda$ plane. 
The grey areas are the location of the 1 and 2-$\sigma$
bands (respectively darker and lighter bands)
allowed by a measured skewness that would be obtained
with either $\Omega_0=0.3$ (left bands) or $\Omega_0=1$ (right bands).
The solid straight lines corresponds to a zero curvature universe, and
the dot-dashed lines to a fixed acceleration parameter, $q_0$. 
The panels correspond to survey
of either $5\times5$ (top) or $10\times10$ degrees (bottom).
}
\end{figure}

In the context of the rapid changes occurring in observational 
cosmology it is worth stressing that weak lensing surveys offer
precious complementary views of our Universe and unique tools to probe {\it directly}
the dark matter and to compare with the light distribution at any scales. 
The perspective of determining the projected power spectrum
independently of biases is indeed attractive. Moreover
the possibility of determining $\Omega_0$ in a way which 
is independent on the power spectrum, and independent on all
the methods that have been suggested so far,  
is also extremely precious. 
We remind that this determination
relies only on dynamical effects assuming that the
large-scale structures originate from Gaussian initial conditions
through gravitational instabilities. In Fig. \ref{OmLambda.ps} examples
of constraints in the ($\Omega,\lambda$) 
plane with weak lensing as describes in
this work (shaded areas) are presented\footnote{The $\lambda$
dependence is taken from the theoretical results given by BvWM.}
together with the location of the major constraints that are expected
to be brought either with CDM experiments, (constant curvature
density, solid lines), or from Type-Ia supernovae  
measurements (the other straight lines, describing constant $q_0$ values).

Future CMB experiments can determine the cosmological
parameters with a remarkable accuracy, but only when some
prior is put on the shape of the initial power spectrum. The
curvature will probably be determined with a good accuracy in the near 
future from the position of the first Doppler peak on
the $C_l$ curves. But it appears that 
it can be very difficult to disentangle $\Omega_0$ from $\lambda$
(see for instance Zaldariagga, Spergel \& Seljak 1997) from the mere
temperature (and polarization) fluctuations. This might be only possible
from a very detailed analyses and the power spectra which require
not only a good understanding of the possible systematics that
may affect the measurements, but also specific hypothesis on the
regularity of the primordial power spectrum. The requirements
to measure $\Omega_0$ from weak lensing survey are less strict
provided the instrumental systematics can be controlled correctly.

Moreover weak lensing surveys are able to constraint
$\lambda$ if it is possible to select efficiently 
several populations of sources (see Villumsen 1996, BvWM
for clues of such possibilities). We thus think that 
weak lensing survey will be a major mean for probing
the global cosmological parameters. It is complementary to the CMB
experiments and indispensable for breaking the parameter
degeneracy. %FB 

The feasibility of the weak lensing by large scale structure program and
The full scientific exploitation of the weak lensing effect is
reduced to the capability to control instrumental
systematics, where there are still open crucial points.
The most dedicated problem comes probably from the need of a 
PSF correction able to avoid artificial large scale
coherent alignments of galaxies. Related to this problem
is the pixelisation effect. Galaxies shapes are indeed determined
from a finite number of pixels with the possible introduction
of errors and biased estimation on the shapes.

These effects can all be investigated independently. Taking 
several images of the same portion of the sky with a shifted
position for the camera, or with different cameras
would definitely test the robustness of the observed distortion maps.
Moreover we have numerous statistical tests at our disposal that
can be done on the maps: comparison of the 2-point correlation function
of the distortion with the one of the shear field or with
the one of the $\kappa$ field, or even by the investigation
of quantities that are a priori sensitive to the systematics 
in a different manner, such as the correlation of the orientations
of the distortion field. That all these quantities have related properties
is somehow due to the fact  that in the thin lens approximation
the $\gamma$ field should be curl-free (see for example Luppino \& Kaiser
1997 for a used of these tests).
This is true only when lens-lens couplings are neglected
but it should be possible to investigate those
effects in numerical simulations.

{
\acknowledgements It is a pleasure to thanks S. Colombi for his useful
suggestion to use a second order Lagrangian dynamics, B. Fort, P. Schneider,
B. Jain and U. Seljak for fruitful discussions. We are very greatfull to
the referee S. Seitz for her extremely detailed report with very
useful remarks and suggestions. This research has been conducted under the auspices of a
European 
TMR network programme made possible via generous financial support 
from the European Commission. This
work was also supported by the Programme National de Cosmologie and the
``Sonderforschungsbereich 375-95 f\"ur Astro-Teilchenphysik'' der
Deutschen Forschungsgemeischaft. L.V.W. and F.B. thank IAP, where a fair
fraction of this work has been done, for hospitality.
}

\section*{Appendix A: Second order Lagrangian dynamics}

In order to investigate the statistical properties of the reconstructed mass
field that contains a significant amount
of non-linearities, a non-linear evolution model of
large-scale structures is required. Second order Lagrangian perturbation
theory has the advantage to be extremely fast to compute, and rather
accurate compared to N-body simulations. It gives correct values
for the skewness $S_3$ and we expect it to gives correct estimation
of the cosmic variance 
(see Munshi et al. 1994, Bernardeau et al. 1994, Bouchet et al. 1995). 

Actually we do not even need to do 3D simulations. At the level
of perturbation theory it is equivalent to perform 2D
second order Lagrangian evolutions of the structures on 
an initial linear map of the projected mass fluctuations.

\subsection*{Construction of the initial linear map}

The local convergence map is given by
\ba
\kappa(\varphig)=&-
\displaystyle{{3\over 2}\ \Omega_0\ \int_0^{\chi_H}\d\chi_s\ n(\chi_s)\ 
\int_0^{\chi_s}\d\chi}\nonumber\\
&\displaystyle{
{\Dc_K(\chi_s-\chi)\ 
\Dc_K(\chi)\over \Dc_K(\chi_s)}\ {\delta({\cal D}_K(\chi)\varphig,\chi)\over a}},
\ea
where $\chi_H$ is the horizon distance, $\Omega_0$ the density 
parameter, $a$ the expansion factor and
$n(\chi_s)$ the number density of sources as a function
of the distance $\chi$.
For clarity we introduce the lens efficiency function
\be
w(\chi)={3\over 2}\ \Omega_0\ \int_{\chi}^{\chi_H}\d\chi_s\ n(\chi_s)\ 
{\Dc_K(\chi_s-\chi)\ 
\Dc_K(\chi)\over a\,\Dc_K(\chi_s)},
\ee
so that the local convergence is simply given by
the integral over the line of sight of,
\be
\kappa(\varphig)=-\int_0^{\chi_H}\d\chi\ w(\chi)\ \delta({\cal D}_K(\chi)\varphig,\chi).
\ee
The projected density contrast would be,
\be
\delta_{2D}={\kappa\over\omb}\ \ {\rm with}\ \ 
\omb=\int_0^{\chi_H}\d\chi\ w(\chi).
\ee
It is important to keep in mind that the local convergence is related
to the actual density contrast with a constant that depends
on the cosmological parameters.
In the following we will assume that all sources are at the same
redshift (it is not realistic but of no consequences
on our results here).

In the linear approximation the $\kappa$ field is expected to be
a 2D Gaussian field, characterized by a power spectrum, $P_{\kappa}$.
with (see Kaiser 1992, 1996, and SvWJK)\footnote{Taking $H_0=c=1$.},
\begin{equation}
P_{\kappa}(\rmk)=\int_0^{\chi_H} \d\chi\ w(\chi)^2\ 
P_{\delta}\left({\rmk\over \Dc_K(\chi)},\chi\right),
\end{equation}
as a result of the relation between $\kappa$ and
the 3D density contrasts.
The initial conditions for $\kappa$ are therefore generated by a 2D Gaussian
field in Fourier space where the complex random variables verify
\begin{equation}
\langle \tilde\kappa(\kg)\tilde\kappa^{\star}(\kg')\rangle=
(2\pi)^2\delta_D(\kg-\kg')P_{\kappa}(\rmk),
\end{equation}

\subsection*{The 2D second order Lagrangian dynamics}

In order to introduce a significant amount of nonlinearities in the
maps, we apply the 2D second order Lagrangian dynamics to the
projected density, $\delta_{2D}$.
Following the notations of Bouchet et al. (1992), transposed in the 2D case,
let us write the 2D Eulerian coordinates 
$\thetag(\qg)$ as a perturbation series over the 2D displacement field $\Dg$,
\begin{equation}
\thetag(\qg)=\qg+\epsilon\Dg^{(1)}(\qg)+\epsilon^2\Dg^{(2)}(\qg)+
o(\epsilon^3),
\end{equation}
where $\qg$ is the angular Lagrangian coordinate 
and $\epsilon$ a small dimension-less
parameter. The first order term reduces to the Zel'dovich
(Zel'dovich 1970) approximation.
The divergence of the second order displacement field can be written in terms
of the first order solutions,
\be
\Dg_{i,i}^{(2)}(\qg)={3\over 7}
\left[\Dg_{1,1}^{(1)}\Dg_{2,2}^{(1)}-\left(\Dg_{1,2}^{(1)}\right)^2\right].
\ee
This result is exact for an Einstein-de Sitter Universe. 
The values of the coefficient $3/7$ is only slightly altered (about 1\%) for
other cosmological models (Bouchet et al. 1992, Bernardeau 1994)
and in the following we did not take into account this dependence.
Once the second order displacement field has been computed,
the local 2D density contrast can then be written in terms
of the Jacobian of the transform between the Lagrangian 
coordinates and the Eulerian coordinates,
\begin{equation}
\delta_{2D}(\thetag(\qg))={1\over J(\qg)}
\end{equation}
where $J=\left|\partial\theta_i/\partial q_j\right|$. The 
local convergence is then given by
\be
\kappa(\thetag(\qg))=\omb\ \delta_{2D}(\thetag(\qg)).
\ee
The linear density maps are built on a regular grid from 
random modes following a given power spectrum. The different
quantities up to the Jacobian are generated on this same grid by successive
Fast Fourier Transforms.
There is then a technical difficulty to solve in order to get the resulting
values of $\kappa$ on a regular grid. This is done via a local
triangulation and an interpolation of the values (from standard IDL packages). 
Note that the continuity equation provides us with
$\kappa$ and not with the projected potentiel. The latter
will have to be computed from a subsequent Fourier transform
(see Appendix B). %FB
The amplitude of the fluctuations is such that the displacement field
does not induce shell crossings\footnote{However, in the open cases
it very rarely happened that $\kappa$ reached unrealistic
large values. The amplitude of $\kappa$ has therefore
been arbitrarily limited to 0.1. At most a handful of pixels
have been affected by this cut-off.}. Finally, 
bands of sufficient width along the edges were cut out to avoid
edge effects induced by the displacement filed.

\subsection*{The skewness with this approximate dynamics}

The skewness of the projected density is given by the skewness
of the 2D dynamics (Munshi et al. 1997), that is,
\be
s_3^{\rm density}={36\over 7}+{3\over2}{\d\log\sigma^2(\theta)\over \d\log\theta}.
\ee
The skewness for the convergence would then simply be
\be
s_3^{\rm convergence}={1\over \omb}\left(
{36\over 7}+{3\over2}{\d\log\sigma^2(\theta)\over \d\log\theta}\right).
\label{s3_2d}
\ee
This result has to be compared with the results obtained in
BvWM. Their general formula (67) contains some extra geometrical factors
of order unity (coming from a different averaging procedure along
the line of sight). Eq. (\ref{s3_2d}) actually corresponds to the
approximate form of Eq. (75) of BvWM.

\subsection*{Shapes and normalizations of the power spectra}

The 3D power spectrum given by Baugh \& Gazta\~ naga (1996) (BG spectrum)
is used in most of our simulations,
\begin{equation}
P(\rmk)\propto {k\over \left[1+(k/k_c)^2\right]^{3/2}},
\label{BGpower}
\end{equation}
where $k_c=0.05~h_{100}{\rm Mpc}^{-1}$. 
In a series of simulations we compared this model also with the
standard CDM spectrum.

In most cases the fluctuations are normalized according to the convergence
field $\sigma_{\kappa}=Cste$. The value of $\sigma_8$\
that has been chosen thus depends on $\Omega_0$ in such
a way that it is equal to $0.6$ for a flat Universe
(following the normalization inferred from galaxy cluster counts (Eke et al.
1996, Oukbir \& Blanchard 1997). As a result we take,
\be
\sigma_8=0.6\ {\omb(\Omega_0=1)\over\omb(\Omega_0)}
\ee
It implies that the value of $\sigma_8$ grows for low values of $\Omega$.
Note that since $\omb\approx\Omega^{-0.8}$, this growth is only
slightly more important in our case than for the galaxy cluster counts
(the exponent would be about 0.5 to 0.6).

\section*{Appendix B}

This appendix gives the details of the reconstruction algorithm,
how the $g$ map is generated from an initial $\kappa$ map, and
the noise model.

\subsection*{Reconstruction algorithm}

The lensing quantities of interest
are defined in the Eqs.(\ref{lens_def}), and the
observable is the reduced shear $g=\gamma/(1-\kappa)$. The reconstruction
problem is how to infer the $\kappa$ map from an observed ellipticity
field $\epsilon_{obs}$, knowing that $\langle\epsilon_{obs}\rangle=g$ is
an unbiased estimate of the reduced shear?
Note that this problem is under-constrained since a change in the potential
$\psi$ of the form,

\begin{equation}
\psi\rightarrow \lambda\psi+{(1-\lambda)\over 2}\left| \thetag\right|^2,
\end{equation}
where $\lambda$ is a constant, leaves the reduced shear $g$ invariant, but
will transform the convergence as $\kappa\rightarrow
\lambda\kappa+1-\lambda$ (see Seitz et al. 1998).
This is the so called mass-sheet degeneracy. Thus in order to get a realistic
convergence map, $\lambda$ has to be determined by forcing $\bar \kappa=0$
at the survey scale. The estimate of $\kappa$ is obtain by a
non-parametric least $\chi^2$ method (Bartelmann et al. 1995). The
image is sampled on a $N\times N$ grid, and the potential
on a $(N+2)^2$ grid. Starting from a guess on $\psi$,
a guess on the reduced shear is obtained and
the following function is minimized with respect to $\psi$,

\begin{equation}
\chi^2=\sum_{ij}\left\vert g(\psi)-g\right\vert^2.
\label{chi2_g}
\end{equation}
The finite difference schemes which are used in order to calculate the second
derivatives of the potential at pixel $(i,j)$ are,

\begin{eqnarray}
\psi_{11}(i,j)&=&\left[\psi(i+2,j)+\psi(i-2,j)-2\psi(i,j)\right]/4\Delta^2\nonumber\\
\psi_{22}(i,j)&=&\left[\psi(i,j+2)+\psi(i,j-2)-2\psi(i,j)\right]/4\Delta^2\nonumber\\
\psi_{12}(i,j)&=&[\psi(i+1,j+1)-\psi(i+1,j-1)\nonumber\\
&&-\psi(i-1,j+1)+\psi(i-1,j-1)]/4\Delta^2,
\label{diff_sheme}
\end{eqnarray}
where $\Delta$ is the pixel size.
We found that these schemes give the best regularization
at small scales and avoid the usual high frequency oscillations in crude
reconstruction schemes. At the edges of the field the shift of
2 pixels in Eqs. (\ref{diff_sheme}) must be only one pixel in the direction
perpendicular to the border since the potential
is sampled on a $(N+2)^2$ grid only. This has the consequence to slightly
increase the noise at the boundaries but does not produce any bias.
Once Eq. (\ref{chi2_g}) is minimized, the $\kappa$ map is found, and then
the condition $\bar\kappa=0$ is imposed on it.

\subsection*{Construction of the initial $g$ map}

A problem that we have to solve in this work is how to get the
shear pattern of a projected mass distribution? Namely from the simulated
$\kappa$ we want to get the corresponding distortion map, put a noise
on it, and reconstruct $\kappa$.
The construction of a distortion map from a convergence map is
unfortunately an under-constrained problem again.
For example any transformation of the potential
$\psi\rightarrow\psi+\bar\psi$ with $\Delta\bar\psi=0$ leaves the convergence
unchanged, but may change the shear (such a solution is for example
$\theta_x^2-\theta_y^2+\theta_x\theta_y$). A peculiar solution for the
potential $\psi$ can be obtained by minimizing the $\chi^2$ function:
\begin{equation}
\chi^2={\displaystyle \sum_{ij}} \left(\kappa-\kappa_{\rm guess}\right)^2.
\end{equation}
The resulting reduced shear is then $g+\bar g$, where $\bar g$ is the
unphysical solution given by $\bar\psi$, but it is possible to reconstruct
the convergence using Eq.(\ref{chi2_g}), since $\kappa$ is unchanged by the
presence of the term $\bar g$. Unfortunately it is no longer the case when
the noise is included via Eq.(\ref{elli_single}), because $\bar g$ explicitly
comes in the denominator. Fortunately, the role of the denominator in 
Eq.(\ref{elli_single}) is weak (this is a one percent effect on each galaxy
if we take $g\simeq\bar g\simeq 0.01$ and $\epsilon^{(s)}\simeq 0.1$), this
is in particular one of the reasons why the weak lensing approximation works
so well. We thus did
not try to correct for the presence of a spurious contribution $\bar g$ 
in all the calculations, since it gives a negligible
contribution to the signal (in terms of power spectrum and moments).

\subsection*{Noise generation}
Once the true $g$ maps are obtained the noise is introduced in a realistic
way, using Eq.\ref{elli_single}: a sample of background galaxies with
random intrinsic orientations is sheared, from which the $\kappa$ map is
reconstructed. The galaxies are observed in a grid of {\it superpixels} of
$2.5$ arcmin size (the minimum size of the $\kappa$ map simulations), in which
the number of galaxies $N_i$ per superpixel $i$ is known. In principle this
number suffers of the
amplification bias (depending on the line of sight matter quantity), but this
effect is neglected here. $N_i$
follows a Gaussian distribution of mean $N_p$ and variance $\sqrt{N_p}$.
Each pixel $i$ of the image gives a local estimate of the reduced shear 
from the $N_i$ galaxies, each
of them having an intrinsic ellipticity $\epsilon_j^{(s)}$ which 
contributes  as a source of noise for the shear signal.
The distribution of $\epsilon_j^{(s)}$ is a truncated normalized Gaussian
defined over the range $[0,1]$,

\begin{equation}
p(\epsilon_j^{(s)}) \propto \exp\left(-\left({\epsilon_j^{(s)}\over
\sigma_\epsilon}\right)^2\right),
\label{elli_def}
\end{equation}

where we choose $\sigma_\epsilon=0.12$. (which corresponds to a typical
axis ratio of $0.8$).

>From Eq. (\ref{elli_single}), the observed mean ellipticity $\bar\epsilon_i$
of a number $N_i$ of galaxies in the image plane is given by

\begin{equation}
\bar\epsilon_i={1\over N_i}{\displaystyle \sum_{j=1}^{N_i}} {\epsilon_j^{(s)} -g\over 1-g^\star\epsilon_j^{(s)}}.
\label{sum_elli}
\end{equation}
Where $\bar\epsilon_i$ is an unbiased estimate of the reduced shear
$g$ in the superpixel $i$ (Schramm \& Kayser 1995, Schneider \& Seitz 1995).
A realistic estimate of $N_p$
depends on the observational context, the telescope used, the optical filter,
the atmospheric conditions. By choosing $N_p=30~$gal/arcmin$^2$ or
$50~$gal/arcmin$^2$ we adopted a conservative and reasonable assumption about
the telescope time: the former is accessible in the I-band with 1.5 hour
integration at CFHT, while the later is accessible for 4 hours in the same
conditions.

\end{document}